# Electron-phonon and phonon-phonon anharmonic interactions in 2H-MoX$_2$ (X=S, Te): A comprehensive Resonant Raman study

Suvodeep Paul, Saheb Karak, Annie Mathew, Ankita Ram, and Surajit Saha*

*Department of Physics, Indian Institute of Science Education and Research Bhopal, Bhopal 462066, India*

(* surajit@iiserb.ac.in)

**Abstract**

Transition metal dichalcogenides (TMDs) are layered materials which show excellent potential for nanoelectronic and optoelectronic applications. However, as many of the exciting features of these materials are controlled by the anharmonic effects, a proper understanding of the phonon properties and anharmonicity associated with these materials is essential for the proposed applications to be realized. We present here a comprehensive study on the phonon properties of two different TMDs; *viz.*, MoS$_2$ and MoTe$_2$, as a function of temperature, laser excitations, and polarization, as well as flake-thickness. Our study includes the measurement of anharmonicity in the first-order and higher order Raman scattering processes. The variations in anharmonicity with the thickness of MoS$_2$ and MoTe$_2$ have been explained in terms of their phonon symmetries, electron-phonon coupling, and phonon-phonon interactions. Further, the effect of the underlying substrate on the anharmonic properties of the in-plane and out-of-plane phonons has also been estimated from the Raman measurements, thus elucidating the intrinsic phonon properties of the 2D layered materials.

**Introduction**

The advancement of technology is closely associated with the miniaturization of devices. While researchers and engineers have been able to successfully miniaturize silicon-based electronic devices till early 2000s following the well-known Moore's Law [1], the feature size of silicon seems to have approached the minimum feasible limit [2]. Among the various techniques proposed to surpass the bottle-neck, is the prediction to use a thinner channel. Therefore, scientists have been interested in a new generation of electronic devices based on lower dimensional materials. The discovery of graphene [3] and its exciting properties like charge mobility [4], flexibility [5], nearly wavelength-independent optical transparency [6] etc. appeared very promising for both nanoscale electronic [1,7,8] and optoelectronic device [7] applications. However, the absence of a bandgap in graphene leads to obvious limitations. But it did not take long to find alternative two-dimensional (2D) materials having semiconducting nature. The transition metal dichalcogenides (TMDs) [9] are layered materials which are semiconducting in their 2H phases and are, therefore, very relevant for nanoscale device applications [10]. Further, the discovery of the transition from indirect to direct band gap upon thinning down from bulk to monolayer in TMDs [11,12] showed further promise in optoelectronic applications. Molybdenum disulphide (MoS$_2$) is the most stable among the TMDs and has been widely explored over the past decade. While monolayer MoS$_2$ shows a bandgap of ~ 1.8 eV [11], the monolayer molybdenum ditelluride (MoTe$_2$) has a bandgap of ~ 1.1 eV [12,13], which coincidentally is comparable to that of silicon (1.1 eV) with the additional benefit of being a direct band gap semiconductor. Efficient functioning of nanoscale devices requires engineering of the thermal properties to ensure proper heat dissipation. Naturally, it is of utmost importance to understand the structural and phononic properties of these 2D materials which play an important role in their electronic, optical, and thermal properties.

Raman spectroscopy is one of the most effective characterizing tools for 2D materials where X-ray diffraction technique is not much useful because of the absence of the third dimension in these materials [14]. Raman response of 2D materials shows excellent dependence on their flake-thickness and can be reliably used to fetch the thickness information of the 2D channels [15-18]. While most of the available studies report the first-order Raman processes which probe only the $\Gamma$-point phonons, a thorough analysis of the phonons in a material involves the study of the higher-order modes as well which can probe phonons from the zone boundaries and other parts of the Brillouin zone. Second-order Raman scattering effectively refers to the processes that involve a change in the phonon occupation number by 2 [19]. It may involve creation of two phonons, destruction of two phonons, or a simultaneous creation of one phonon and destruction of the other. In the first two cases, the second-order mode can be identified in the Stokes and Anti-Stokes side of the spectrum, respectively. However, in the third case, the appearance in the Stokes side depends on whether the energy of the phonon created is greater than the destroyed phonon and vice versa. The second-order Raman scattering process may be classified into line spectrum and continuous spectrum. The line spectrum occurs due to a summation or difference mode resulting from two successive allowed first-order Raman scatterings. On the other hand, the continuous spectrum results when the photon interacts with two phonons in the same event. In these processes, phonons effectively from the entire Brillouin zone are allowed with the requirement that the momentum conservation rules are satisfied. Therefore, the continuous second-order Raman scattering spectrum is proportional to the density of lattice states [20]. The requirement of conservation of momentum may also be mediated by the defects present in the crystal. Therefore, these processes can be effectively used to reveal many important aspects of the material investigated, *viz.*, mapping out the phonon density of states, quantifying the defects and dopings, and so on that play very vital role in the properties of the material. It has been reported [21] that these second-order processes have a strong dependence on the layer-thickness of the TMDs, thereby, allowing the determination of thickness of TMDs based on these measurements. Unfortunately, these (higher-order) processes are generally very weak due to low scattering cross-section and, hence, are observed when the photons couple with an electronic transition, resulting in an enhancement in the Raman response of the material, also known as Resonant Raman process. On the other hand, the photoluminescence spectra of semiconducting TMDs, that arise due to electronic transitions, are characterized by various excitonic features (A, B, and C) and trions in the visible and ultraviolet regions of the electromagnetic spectrum [22,23]. Recently, a few groups have reported [24-28] the observation of resonant Raman processes in these TMDs using appropriate excitation wavelengths that resonate with the excitonic transitions. Importantly, 2H-MoS$_2$ shows excitonic transitions at the *K*-point of the Brillouin zone [22]. On the other hand, the band structure of 2H-MoTe$_2$ shows a saddle point at the *M*-point which gives rise to a Van Hove singularity [28]. Therefore, probing the second-order Raman processes in these materials is possible using the suitable laser excitation sources, and this can be an effective way to study various important properties associated with the phonons.

The phonon properties have historically been studied under the harmonic approximation, wherein the phonons are considered to be non-interacting. This is a fairly accurate approximation at low temperatures when the smallness factor, η, which is defined as the ratio of the atomic displacements to the interatomic distances, is small [29]. However, at finite temperatures, as η increases, the phonon interactions with other quasiparticles (e.g., phonons, conduction electrons, spins, excitons [30] etc.) become important. Phenomena which may not be adequately explained by considering the ideal phonon gas picture are termed as anharmonic. Anharmonicity can strongly influence the thermodyanamic quantities like thermal expansion, thermal conductivity, specific heat etc. [29]. Unlike harmonic phonons, anharmonicity causes phonons to have a finite lifetime and also reduced self-energy. Anharmonicity also strongly influences the thermal transport, electronic, and opto-electronic properties of the materials. Therefore, investigation of the temperature evolution of the phonons is important to understand the anharmonicity, which in turn is essential to estimate many important physical properties of the material and can provide information about the various quantum interactions and correlations involving the phonons. Recent reports on the temperature-dependent Raman studies of TMDs [31-37] have attributed the thermal behaviour of the phonons to thermal volumetric expansions of the crystal (quasi-harmonic effects) and the phonon-phonon anharmonic interactions, involving three-phonon (cubic anharmonicity) and four-phonon (quartic anharmonicity) processes. The anharmonicity associated with these 2D materials is very sensitive to the flake-thickness and the temperature range under investigation. However, to the best of our knowledge, there has never been a comprehensive study on these materials, investigating the various second-order Raman processes, which are normally very weak, probing their dependence on the layer-thickness as well as temperature. It is also important to note that 2D materials are generally supported on a substrate for most practical purposes. It has been recently reported that the thermal properties of 2D materials may be strongly affected by the underlying substrate [38,39]. The thermal expansion coefficients of the substrate and the 2D flake generally show a mismatch that monotonically increases as a function of temperature giving rise to a tensile or compressive biaxial strain on the 2D flake. Therefore, the substrate-supported 2D flakes can appreciably overestimate or underestimate the anharmonicity of the investigated 2D layers. It is absolutely essential to estimate the effect of the biaxial strain induced by the substrate in the thermal studies.

Another important application of Raman measurements is the use of polarization dependence to determine the crystal orientation. Further, the polarization-dependent Raman studies provide in-depth knowledge about the phonon symmetries and their assignments. There are reports of polarization angle-dependent Raman studies performed on $MoS_2$ and $MoTe_2$ [40-42], however, the studies are restricted to their first-order modes alone. To the best of our knowledge, the study of the polarization angle dependence of the second-order modes and the forbidden processes which may be probed through resonant Raman processes have not yet been explored.

In this work, we have studied the various first-order and second-order Raman responses of $MoS_2$ and $MoTe_2$ as a function of temperature and layer-thickness with appropriate laser excitation sources suitable for resonant activities as well as the polarization. Our comprehensive study reveals variable extents of anharmonic behaviour for the various modes. While the first-order modes show temperature coefficients comparable to previous reports, we have observed an intriguing dependence on the flake-thickness which may be attributed to a competition between the electron-phonon interactions and the anharmonic phonon-phonon interactions varying with the layer-thickness. Further, in order to accurately estimate the effects of anharmonicity in the investigated materials, we have estimated the substrate effect on the temperature-dependence of the $E_{2g}$ and $A_{1g}$ phonons in monolayer $MoS_2$. Finally, we have performed extensive polarization angle resolved studies and observed strong polarization angle dependence of the second-order Raman processes. We have also observed interesting departure of Raman selection rules for the in-plane modes at resonant excitation energies.

**Experimental details**

Flakes of variable thickness of $2H-MoS_2$ and $2H-MoTe_2$ were obtained by micromechanical exfoliation technique using scotch tape. The exfoliated flakes were then transferred on to a silicon substrate with a coating of ~ 300 nm thick $SiO_2$ on top. The thicknesses of the obtained flakes were confirmed using an Agilent 5500 atomic force microscope (AFM) as well as Raman spectroscopy. All the Raman measurements were performed using a Horiba JY LabRam HR Evolution Raman spectrometer in the back-scattering geometry and the detection was done by an air-cooled CCD detector. The $MoS_2$ flakes were excited by three different laser lines: (i) 532 nm (Nd:YAG-diode laser), (ii) 633 nm (He-Ne gas laser), and (iii) 325 nm (He-Cd gas laser). Experiments on $MoTe_2$ flakes were performed using 532 nm excitation only because the Raman spectra obtained using the 633 nm and 325 nm laser lines did not show any appreciable resonant enhancement. The measurements with the 532 nm and 633 nm laser lines were performed using a 50× objective (numerical aperture = 0.5) and a grating of 1800 grooves/mm, while the measurements with the 325 nm laser were done using a 15× objective (numerical aperture = 0.3) and a grating of 2400 grooves/mm. Temperature-dependent Raman measurements were performed using a Linkam stage (model no. HFS600E-PB4) by varying the temperature from 80 to 300 K, using low laser power (~ 1 mW) to avoid local heating effects due to the laser beam. Polarization angle-dependent studies are performed using half waveplates compatible with visible and UV wavelengths and an anlyzer for the scattered beam. A detailed description of the arrangements and configurations are provided in Supplemental material (Supplemental note 6) [43].

**Results and discussion**

## I. Determination of the thickness of MoS$_2$ and MoTe$_2$

Figure 1(a) shows the structure of 2H-MX$_2$ (M = Mo, X = S, Te). The unit cell consists of two hexagonal units of MX$_2$ conserving the inversion symmetry. Each monolayer of MX$_2$ is essentially constituted of a layer of M atoms sandwiched between two layers of X atoms. These X-M-X units then form vertical stacks with weak van-der Waals type interlayer interaction in multilayer flakes. Figures 1(b) and (e) show the optical images of bilayer MoS$_2$ and MoTe$_2$ flakes, respectively. Optical images of all other investigated flakes are shown in Supplemental figures S1 and S2 of Supplemental material [43]. The thicknesses of the various investigated flakes were measured using atomic force microscope (AFM) operated in the contact mode. Figures 1(c,d) and (f,g) show the AFM images and the corresponding topographic profiles of the bilayer MoS$_2$ and MoTe$_2$ flakes, respectively. The AFM images for all other investigated flakes are included in Supplemental figures S1 and S2 of Supplemental material [43]. The flake-thicknesses are further confirmed by the low-frequency interlayer Raman modes (breathing and shear modes) obtained at room temperature (see Supplemental note 1 and Supplemental figures S1 and S2 in Supplemental material [43]).

## II. Resonance enhancement of the Raman modes and their assignments

The bulk 2H phase of MX$_2$ (M=Mo, X=S, Te) belongs to the $D_{6h}$ point group and each unit cell contains 2 'M' atoms and 4 'X' atoms. The corresponding phonon modes are represented by the irreducible representation $\Gamma_{\text{bulk}} = A_{1g} + 2A_{2u} + 2B_{2g} + B_{1u} + E_{1g} + 2E_{1u} + 2E_{2g} + E_{2u}$. Out of the members of the irreducible representation, one of the two doubly degenerate $E_{2g}$ modes and one of the two optically inactive $B_{2g}$ modes represent the interlayer shearing and layer breathing modes, respectively [44]. All the other modes represent the intralayer vibrations. As observed in 2H-MoS$_2$, while the intralayer $E_{2g}$ and $A_{1g}$ modes are Raman-active first order modes, the other intralayer modes are forbidden by Raman selection rules. On the other hand, the few-layer thick flakes of MX$_2$ may be categorized into even (point group: $D_{3d}$) and odd (point group: $D_{3h}$) number of layers and the corresponding irreducible representations are given as $\Gamma_{\text{even}} = \frac{3N}{2}(A_{1g} + A_{2u} + E_g + E_u)$ and $\Gamma_{\text{odd}} = \frac{3N-1}{2}(A_1' + E'') + \frac{3N+1}{2}(A_2'' + E')$, respectively, where N represents the number of layers in the flake [44]. For even N, the interlayer shearing (layer breathing) modes are either Raman-active $E_g$ ($A_{1g}$) modes or IR-active $E_u$ ($A_{2u}$) modes. However, for odd N, the interlayer shearing modes are all Raman-active $E''$ and $E'$ (also IR-active), while the layer breathing modes are either Raman-active $A_1'$ or IR-active $A_2''$. The other modes constitute the intralayer vibrations and may be Raman-active first order modes ($E_g$, $E''$, $E'$, $A_{1g}$, $A_1'$) or silent ($E_u$, $A_{2u}$, $A_2''$) [44]. The forbidden modes may however be observed due to breaking of selection rules in case of Resonant Raman processes as explained below. In the following discussions, we will refer to the different modes by the symmetry of the corresponding mode in the bulk crystal, according to the convention followed in most of the available literature.

Figures 2(a) and (b) show the room temperature Raman spectra of 8L-MoS$_2$ and 2L-MoTe$_2$, respectively. We have performed Raman studies on exfoliated MoS$_2$ flakes of various thicknesses (1L, 2L, 3L, 8L, and bulk) with different excitations (see Supplemental note 2 in Supplemental material [43]). In Figure 2(a), the room temperature Raman spectrum of 8L-MoS$_2$ is shown using three different laser excitations: 325, 532, and 633 nm. Importantly, when excited with 633 nm (1.96 eV) laser, the photoluminescence arising from the underlying exciton features in the atomically thin flakes of MoS$_2$ is so strong (due to transition from an indirect to direct band-gap in monolayer and resonance with the laser excitation line) that the second-order Raman modes are hardly visible for analysis. Therefore, 8L-MoS$_2$ has been shown here where the photoluminescence gets considerably suppressed thus bringing out the Raman modes very clearly for analysis using all the different laser excitation sources. The different first-order and second-order Raman modes observed in 8L-MoS$_2$ flakes using the non-resonant (532 nm) and resonant (325 nm and 633 nm) excitation lasers are labelled as F$_1$ to F$_4$ (first-order modes) and ω$_1$ to ω$_{15}$ (combinational modes or second-order modes). Table 1 lists the assignments of all the modes based on previous theoretical and experimental studies [24-28]. It can be observed that the modes show variable enhancements when excited with different excitation wavelengths. The first-order modes $F_1$ and $F_4$ and the combinational modes $\omega_{13}$ to $\omega_{15}$ show strong enhancement with 325 nm excitation. The $F_1$ mode (at ~ 286 cm$^{-1}$) is a $\Gamma$-point phonon involving $E_{1g}$ symmetry ($E''$ for 1L-MoS$_2$) which is forbidden for monolayer and bulk MoS$_2$ in the back-scattering geometry but is allowed for all the intermediate layer thicknesses [45]. The $F_4$ mode (at ~ 472 cm$^{-1}$) is a $\Gamma$-point phonon of $A_{2u}$ symmetry ($A_2''$ for 1L-MoS$_2$) which is Raman inactive, but it has been observed with 325 nm excitation where resonance with high energy band gap results in a violation of the selection rules [24]. The 325 nm excitation corresponds to transition to the continuum flat band levels located between the $\Gamma$ point and the K point of the Brillouin zone and the nature of the resonance resembles the C excitonic resonance [45]. The enhancement of various modes with 325 nm excitation can be attributed to the non-confined nature of the C excitons as opposed to the A and B excitons which are confined to single layers [45] (See Supplemental Note 2 in Supplemental material [43]). On the other hand, the high frequency combinational modes ($\omega_{13}$ to $\omega_{15}$) are due to triple resonant processes mediated by splitting of the valence band away from the $K$-point [24]. The modes labelled as $\omega_1$ to $\omega_{10}$ are also combinational modes representing processes away from the $\Gamma$-point and show strong enhancement at 633 nm (1.96 eV) excitation. The modes $\omega_2$ and $\omega_4$, which are marked as c and d, respectively, in Table 1, following Lee *et al.* [24], may be assigned as follows: The $\omega_2$ mode (~ 425 cm$^{-1}$) is a combination of forbidden $E_{1u}^2$ phonon and a dispersionless phonon near 38 cm$^{-1}$ (whose origin is still unclear) mediated by the exciton-polariton bands [24] while the $\omega_4$ mode (~ 467 cm$^{-1}$) is a phonon due to $A_{2u}(\Gamma)$ branch away from the $\Gamma$-point [24] or a phonon arising from the combination of $E_{1g}(M)$ and an acoustic phonon at the $M$-point [25,26]. Based on the variable resonance enhancements at different excitation energies, we have probed the various modes in our study using different excitation wavelengths, as mentioned in Table 1. Similar to MoS$_2$, MoTe$_2$ also exhibits first-order and higher-order/combinational modes that resonate with specific laser excitation sources. As shown in Figure 2(b), the bilayer MoTe$_2$ shows first-order Raman modes ($F_1$ to $F_4$ at the $\Gamma$-point), forbidden single-phonon processes ($P_1$ to $P_4$) away from the $\Gamma$-point, and multi-phonon ($\omega_1$ to $\omega_6$) processes. The phonon assignments of the various modes of MoTe$_2$ are enlisted in Table 2. The band structure of MoTe$_2$ exhibits a saddle point at the $M$-point which gives rise to 2D Van Hove singularities with bandgaps of 2.30 eV and 2.07 eV, which in turn give rise to strong resonance enhancement of the second-order Raman modes in MoTe$_2$ when excited with the 532 nm (2.33 eV) and 633 nm (1.96 eV) wavelengths. The Raman spectra for various flakes of MoTe$_2$ were also measured at room temperature using 633 nm excitation (see Supplemental Figure S8 in Supplemental material [43]). Importantly, it was observed that the spectra of MoTe$_2$ due to the 633 nm excitation were very similar to the corresponding spectra obtained using the 532 nm excitation (see Supplemental Figure S7 in Supplemental material [43]) because of the resonance with both the laser excitation sources. Therefore, the thermal studies on the various flakes of MoTe$_2$ were performed using 532 nm excitation. Like in MoS$_2$, the $F_1$ ($E_{1g}$ symmetry) and $F_4$ ($B_{2g}$ symmetry) modes are forbidden in monolayer and bulk MoTe$_2$ (see Supplemental Figures S7 and S8 in Supplemental material [43]). However, the first-order modes $F_2$ and $F_3$ of MoTe$_2$, that are of $A_{1g}$ and $E_{2g}$ symmetries, respectively, are swapped in positions with respect to MoS$_2$. Notably, the single-phonon processes away from the $\Gamma$-point are also forbidden according to the Raman selection rules. However, strong resonance with the band gap at the $M$-point causes the violation of the selection rules, therefore, resulting in the appearance of all these forbidden modes in the Raman spectra. On the contrary, the forbidden modes $P_1$ to $P_4$ become almost invisible for thicker flakes and could be analysed only for the very thin (1L and 2L) flakes. This can be understood as an effect of the transition from a direct to indirect band gap in thicker MoTe$_2$ [12]. Unlike other TMDs, MoTe$_2$ exhibits a direct band gap for both monolayer and bilayer flakes and an indirect band gap for thicker flakes, resulting in a reduction of the exciton photoluminescence yield beyond a thickness of two layers [12]. Naturally, the multilayer flakes show poor resonance, resulting in the disappearance of the forbidden modes.

### III. Anharmonicity of phonons

The thermal behaviour of phonons in MoS$_2$ and MoTe$_2$ has been studied by various groups [31-37]. The phononic behaviour as a function of temperature has been largely attributed to anharmonic effects. The two main contributions to the anharmonicity in the lattice are the volumetric thermal expansion (quasi-harmonic approximation) and the phonon-phonon interactions. As both MoS$_2$ and MoTe$_2$ have been reported to show very low thermal expansion coefficients [46], the phonon-phonon interactions may be considered as the main contributor to the thermal behaviour of phonons [31]. The phonon-phonon interactions may be constituted of the three phonon (cubic) processes, which is essentially considered as the decay of a $\Gamma$-point phonon having energy $\hbar\omega$ into two acoustic phonons of energy $\hbar\omega/2$, having equal and opposite momenta [47]. Similarly, there can be other decay channels for the optical phonons involving four-phonon (quartic) processes. In the cubic anharmonic process, the phonon anharmonicity is manifested by a softening of the mode frequency and broadening of the linewidth with increasing temperature. The phonon frequency and linewidth vary as a function of temperature as per the relations given below [48,49]:

$$\omega(T) = \omega_0 - A\left(1 + \frac{2}{e^{\frac{\hbar\omega_0}{2k_BT}} - 1}\right) \qquad (1)$$

$$\Gamma(T) = \Gamma_0 + B\left(1 + \frac{2}{e^{\frac{\hbar\omega_0}{2k_BT}} - 1}\right) \qquad (2)$$

where $\omega_0$ and $\Gamma_0$ are the frequency and linewidth corresponding to the phonon at absolute zero temperature, while $A$ and $B$ are coupling constants. Though these equations take care of the three phonon processes, consideration of the four-phonon processes as well, that may start to contribute at very high temperatures (above Debye temperature), would result in addition of further terms in equations (1) and (2) [50]. It may be noted here that 2H-MoS$_2$ has been reported to show a Debye temperature in the range of 260 K - 320 K [51]. The temperature dependence of the phonons has been reported to show different behaviours at different temperature ranges. Therefore, for a better understanding, we may consider the behaviour of phonons at three different temperature regimes. It has been reported that at low temperatures (<100 K), the frequency and linewidths of the various Raman modes show no or minimal change as a function of temperature indicating an absence of anharmonicity at these temperatures [47]. At intermediate temperatures between 100 K to 300 K, equations (1) and (2) effectively reduce to linear functions of temperature, whereas at higher temperatures it enters the non-linear regime. Therefore, various reports [31-37] of temperature-dependent Raman studies of MoS$_2$ and MoTe$_2$ have shown a linear temperature dependence of phonon frequencies and linewidths in the temperature range of 100 K to 300 K. This linear temperature-dependence is also captured in the Grüneisen model which follows the equations (3) and (4) for phonon frequency and linewidth, respectively, given below:

$$\omega(T) = \omega_0 + \chi T \qquad (3)$$

$$\Gamma(T) = \Gamma_0 + \xi T \qquad (4)$$

where $\omega_0$ (or $\Gamma_0$) is the phonon frequency (linewidth) extrapolated to 0 K and $\chi$ (or $\xi$) is the first-order temperature coefficient for the phonon frequency (or linewidth) and $T$ is the measured sample temperature. The first-order temperature coefficients $\chi$ and $\xi$ take into account both the intrinsic anharmonic effects i.e., phonon-phonon coupling and the volumetric expansion (quasi-harmonic) effects on the thermal behaviours of the phonons. Based on the anharmonicity theory [48-50], we know that the phonons show a redshift with an increase in temperature. This results in a negative $\chi$-value for all the phonons in absence of other quantum many-body interactions involving various degrees of freedom, like spin and charge, for example. The linewidth (FWHM) of a phonon mode represents the inverse of the lifetime of the corresponding phonon. Furthermore, following the anharmonic behaviour of phonons, the phonon population is expected to increase with temperature which in turn results in a smaller mean free path and a shorter lifetime for the phonons. Therefore, we expect to see an increase in the linewidths as a function of temperature, which would result in a positive $\xi$-value when fitted with the equation (4). In case of second-order Raman scattering, like the first-order processes, the anharmonicity is manifested by a redshift in phonon frequency and a broadening of the linewidth [19]. Ideally, in case of summation modes, the frequency shift of the second-order mode follows the summation of the individual phonon shifts. However, in presence of anharmonic phonon-phonon interactions, the resulting frequency shift is often lesser than the summation of the individual phonon shifts due to the phonon-phonon binding energy. On the other hand, the difference modes exhibit minimal shifts. We may, therefore, infer that all the second-order modes investigated would show similar linear temperature dependences of the phonon frequencies with a much higher first-order temperature coefficient ($\chi$) value in case of summation modes and a very low coefficient value for the difference modes.

**IV. Effect of temperature on phonon frequencies**

Raman spectra obtained at various temperatures were fitted using Lorentzian multi-functions for analysis. The obtained peak positions (phonon frequencies) and linewidths (to be discussed in Section V) for various modes have been studied as a function of temperature. Figures 3 and 4 show the dependence of the various phonon frequencies of 2L-MoS$_2$ (obtained using 532 nm and 325 nm excitations) and 8L-MoS$_2$ (obtained using 633 nm excitation), respectively, as a function of temperature. Figure 5 shows the phonon frequency of 2L-MoTe$_2$ as a function of temperature (the temperature-dependent spectra of MoS$_2$ and MoTe$_2$ are provided in Supplemental figures S10-S13 in Supplemental material [43]). The frequencies of the modes show variable degrees of redshift as a function of temperature, as indicated by their corresponding first-order temperature coefficients ($\chi$'s). It is observed that the $\chi$'s exhibit a wide range of values, depending on the symmetries of the concerned phonons. Similarly, temperature-dependent studies performed on other flakes of MoS$_2$ and MoTe$_2$ having different flake-thicknesses (See Supplemental note 3 of Supplemental material [43]) reveal a systematic variation in $\chi$'s as a function of flake-thickness. The $\chi$'s obtained for various modes of MoS$_2$ and MoTe$_2$ flakes of varying thicknesses are enlisted in Tables 3-5. We will discuss the trends followed by the first-order temperature coefficients for the different modes of MoS$_2$ and MoTe$_2$ based on the following: (i) variation of symmetry of the phonon associated and (ii) the variation of the flake-thickness.

For a better comparison, we have plotted the $\chi$-values of the various first-order and combinational modes of MoS$_2$ and MoTe$_2$ as a function of the flake-thickness in Figure 6. The variations in $\chi$-values observed in Figure 6 are discussed below:

(i) <u>Dependence on the phonon symmetries</u> – The variable temperature dependences of the first-order modes of MoS$_2$ and MoTe$_2$ have a direct connection to the symmetry of the associated phonons. The symmetry of the phonons determines the strength of their interactions with the electrons, which in turn plays an important role in the thermal behaviour of these 2D materials. For MoS$_2$, the F$_3$ mode (having $A_{1g}$ symmetry) shows higher absolute values of $\chi$ as compared to the in-plane modes F$_1$ ($E_{1g}$) and F$_2$ ($E_{2g}$). The first principles calculations by Lazzeri *et al.* [52] suggest that the phonons which strongly couple with electrons tend to exhibit greatest anharmonic frequency shifts. Importantly, it has also been reported by various groups that the $A_{1g}$ (out-of-plane) phonons of MoS$_2$ show a stronger coupling with the electrons compared to the in-plane phonons [32,47,53,54]. This is further established from our Raman measurements (to be discussed in Section V). Therefore, we can conclude that the higher $\chi$-values observed in the case of F$_3$ mode ($A_{1g}$) is because of the stronger coupling of the $A_{1g}$ phonons of MoS$_2$ with the electrons. Similarly, among the first-order modes of MoTe$_2$, the out-of-plane F$_4$ mode (possessing $B_{2g}$ symmetry) displays strongest coupling with the electrons, and hence the highest $\chi$-value. It may be noted that the forbidden modes P$_1$ and P$_2$ of MoTe$_2$, assigned to *TA(M)* and *LA(M)* phonons, respectively, (see Figure 5) show high degree of anharmonicity, giving rise to high values of $\chi$ for 2L-MoTe$_2$ due to strong electron-phonon coupling.

We observe that for any thickness of MoS$_2$ or MoTe$_2$, most of the combinational modes display a higher value of $\chi$ as compared to the first-order modes. The high $\chi$-values of the summation modes are expected as the effective frequency change in a summation mode ideally should be equal to the sum of the frequency changes of the individual phonons. Though, this might not always be true because of the phonon binding energies associated with the combinational modes, but it is expected that the combinational modes should exhibit higher temperature coefficients. It is important to note that the $\chi$-value obtained for the $\omega_2$ mode (Figure 4) is the largest among the combinational modes probed using 633 nm excitation. The large $\chi$-value is attributed to the strong electron-phonon coupling associated with this mode [55]. Further discussions on this mode are provided in Supplemental note 4 of Supplemental material [43]. It is also observed that the $\omega_3$ mode (assigned to 2$LA$ phonons) shows a very high anharmonicity for monolayer MoS$_2$ that may have a significant impact on the thermal properties. This is a very important observation for a better understanding of thermal transport properties of MoS$_2$. Lastly, the difference modes show a negligible change in frequency as a function of temperature, as discussed earlier, and hence the $\omega_2$ mode of MoTe$_2$ (having a phonon assignment of $E_{2g}(M)$-$TA(M)$) shows the lowest $\chi$ values for all thicknesses.

(ii) <u>Dependence on the flake-thickness</u> – As discussed above, the anharmonic frequency shifts associated with phonons depend on their coupling strengths with electrons, which again is dictated by the symmetries of the phonons. Therefore, in this discussion about the thickness dependence of the phonon anharmonicities, we will observe how the anharmonicity associated with the phonons of a particular symmetry tends to vary as a function of thickness. The first-order temperature coefficients of the various modes show strong dependence on the flake-thickness. For MoS$_2$, the first-order modes F$_2$ and F$_3$ of symmetries $E_{2g}$ and $A_{1g}$, respectively, show a decrease in $\chi$ values with increasing flake-thickness due to the additional van-der Waals interactions resulting in an increase in the corresponding force constants [56]. The suppression in the temperature coefficient with increasing flake-thickness is stronger in the F$_3$ mode as compared to F$_2$ because the additional van-der Waals interactions affect the out-of-plane modes more severely. Interestingly, the modes F$_1$ ($E_{1g}$) and F$_4$ ($A_{1g}^2$) show an increase in the temperature coefficient for the bulk flakes with respect to their few-layer counterparts. It must be noted that the modes F$_1$ and F$_4$ appear due to a resonance of the C excitons with the laser excitation source and the C excitons, unlike the A and B excitons, are not confined to individual single layers, but spread over all the layers of MoS$_2$ [45]. Therefore, the C excitons would promote a stronger electron-phonon coupling, resulting in a larger temperature coefficient for F$_1$ and F$_4$ modes in the thicker layers (bulk flakes), as observed in Figure 6. On the other hand, the first-order modes of MoTe$_2$ show lesser dependence on layer thickness. However, like MoS$_2$, the F$_2$ ($A_{1g}$) mode of MoTe$_2$ shows a strong suppression of $\chi$ with increasing layer thickness which can be attributed to the van-der Waals interactions that affect the out-of-plane modes more severely. The $\chi$'s for the combinational modes of both MoS$_2$ and MoTe$_2$ are nearly insensitive to the layer thickness, except for the $\omega_3$ (2$LA$) mode in MoS$_2$. The strong suppression of $\chi$ for the $\omega_3$ mode in MoS$_2$ may be linked to the renormalization of the phonon dispersion and anharmonic phonon scattering rates of the $LA$ phonons of MoS$_2$ with the addition of layers [57]. This can be a fundamentally important factor leading to the strong suppression of the thermal conductivity of MoS$_2$ with an increase in the flake-thickness [57].

## V. Effect of temperature on phonon linewidths

Figures 7(a) and (b) show the linewidths (FWHMs) for the first-order modes of MoS$_2$ and MoTe$_2$, respectively, obtained using 532 nm excitation. As discussed previously, the linewidth corresponds to the inverse of the lifetime of the associated phonon. It may be noted that the phonon lifetimes (and hence, the phonon linewidths) are dependent on both electron-phonon interactions and anharmonic phonon-phonon interactions [58,59]. Due to phonon anharmonicity, the phonon scattering rates increase with increasing temperature, which in turn results in a decrease in the phonon lifetimes (i.e., a broadening of the linewidths) with temperature. In contrast to phonon anharmonicity, the electron-phonon interactions have been reported to cause narrowing of linewidths as a function of temperature [58-60]. A competition between the phonon anharmonicity and the electron-phonon interaction terms decides the effective behaviour of the mode linewidths (phonon lifetimes) as a function of temperature. As depicted in Figure 7, we observe a linearly increasing trend for the linewidths of the various first-order modes of MoS$_2$ and MoTe$_2$. This indicates that the effect of phonon-phonon interactions is dominant in the investigated materials. The FWHMs in Figure 7 are fitted with equation 4 and the corresponding obtained $\xi$-values are enlisted in the Supplemental material [43] (Table ST2). We observe that for all thicknesses of MoS$_2$, F$_3$ ($A_{1g}$) mode shows a higher rate of broadening ($\xi$). Similar to the discussion about phonon frequencies, the $\xi$-values obtained from the mode linewidth vs temperature plots again may be discussed by considering their dependences on the phonon symmetries and the flake-thicknesses. While considering the dependence on the phonon symmetries, we may recall the stronger coupling of the $A_{1g}$ phonon with the electrons [32,47,48], resulting in greater extents of anharmonicity associated with the $A_{1g}$ phonons (as discussed in Section

IV). This explains the higher $\xi$-values exhibited by the $F_3$ mode, indicating greater anharmonicity associated with the $A_{1g}$ phonons. Further, while understanding the thickness dependence, we consider the evolution of the phonons of a particular symmetry with thickness so that all the phonons compared are equally likely to couple with the electrons. It is interesting to observe that for both $MoS_2$ and $MoTe_2$, the temperature susceptibility of the phonon linewidths of the first-order modes decreases with increasing flake-thickness. This trend can be very well explained by considering an increase in the electron population with an increase in the thickness, resulting in an effective increase in electron-phonon interactions with the addition of layers to the flake. It can be observed that the effective trend still shows a positive slope ($\xi$) for the bulk flakes, implying that the phonon anharmonicity is still the dominant term. However, the increase in the electron-phonon coupling with the increase in the flake-thickness is consistent with the decrease in the $\xi$-values with an increase in flake-thickness. Importantly, the $F_3$ mode ($A_{1g}$) of $MoS_2$ again shows a stronger suppression of the $\xi$-value as compared to the $F_2$ mode ($E_{2g}$) (See Figure S17 in Supplemental material [43]). As the $A_{1g}$ phonons show a stronger coupling with the electrons, therefore, an increase in electron-phonon coupling (due to addition of layers) results in the strongest effect on the $A_{1g}$ phonons.

Though the incorporation of the electron-phonon coupling term in the first-order modes provide an accurate explanation of the overall thermal behaviour, the effect of the electron-phonon interactions on the linewidth is only indirect. This means that we do not observe any actual anomalous behaviour in the phonon linewidth trends as a function of temperature, which is expected in presence of electron-phonon coupling, as discussed above. Instead, we observe the effect of the electron-phonon interactions mediated as a decrease in the extent of the anharmonic phonon-phonon interactions with the addition of layers to the flake. However, a direct evidence of the electron-phonon interactions (manifested by anomalous behaviour of the phonon linewidth as a function of temperature) is also observed for the combinational mode $\omega_2$, probed using the 633 nm excitation source (Refer to Supplemental note 4 of the Supplemental material [43]). It is to be noted that this mode has been previously reported to be associated with electron-phonon coupling by Rao *et al.* [55], suggesting the presence of a stronger electron-phonon coupling in bulk flakes of $MoS_2$ as compared to thinner nanoflakes.

### VI. Competition between electron-phonon and anharmonic phonon-phonon interactions

A comparison of the thermal effects on the phonon frequencies and linewidths with varying flake-thickness brings out the contributions of the electron-phonon coupling and the phonon-phonon anharmonic interactions in $MoS_2$ and $MoTe_2$. We observe that the two terms also show mutual dependence because a stronger anharmonicity is observed for the phonons which couple strongly with the electrons. Therefore, the symmetry of the phonons is an important factor in understanding their thermal behaviour. We have also discussed how a competition between the electron-phonon and phonon-phonon interaction terms leads to the effective behaviour of a phonon mode as a function of temperature. This is particularly important in understanding the evolution of the anharmonicity of the phonons of a particular symmetry with the thickness of the flake. We observe that the phonon linewidths exhibit lesser sensitivity to temperature with an increase in the flake-thickness. This could be understood as a result of an increase in the electron-phonon interactions with addition of layers to the flake. Further evidence of electron-phonon coupling is obtained by the anomalous thermal behaviour exhibited by one of the combinational modes, as discussed in Supplemental note 4 (Supplemental material [43]). The anomaly increases with increasing flake-thickness, thereby confirming the presence of higher electron-phonon coupling in thicker flakes. This can also be connected to the decrease in $\chi$-values (and hence a decrease in anharmonicity) with the increase in the flake-thickness (as discussed in Section IV). An overall suppression of the phonon anharmonicity is, therefore, due to an increase in the electron-phonon interactions with the increase in the flake-thickness.

### VII. Effect of the underlying substrate

With the above knowledge of the temperature coefficients for the different modes, we have quantified the thermal effect induced by anharmonicity in $MoS_2$ and $MoTe_2$. However, in all the experiments, the investigated flakes were supported on Si substrates coated with ~ 300 nm thick $SiO_2$ film. For most practical purposes and applications, it is customary to use a substrate, e.g., the conventional dielectric $SiO_2$/Si substrate, to support the 2D flakes. The interaction of the 2D layer with the underlying substrate plays a major role on its thermal behaviour [39,61] and hence on its properties. The effect is particularly strong for atomically thin monolayer flakes. The thermal expansion coefficients (TECs) of the 2D material and the substrate show considerable mismatch as a function of temperature, which in turn gives rise to a biaxial strain on the 2D flakes investigated. The biaxial strain would resist the volumetric expansion of the 2D flake, thereby affecting the thermal evolution of the phonons. In order to investigate the substrate-induced biaxial strain effect on the monolayer $MoS_2$, we have considered a model by Yoon *et al.* [61]. The frequency shift of a phonon as a function of temperature $[\Delta\omega = \omega(T) - \omega_0]$ as seen in equations (1) and (3) should be a property of the 1L-$MoS_2$ alone. But, due to the TEC mismatch with the substrate, the frequency shift may be rewritten as follows [61]:

$$\Delta\omega = \Delta\omega_{actual} + \Delta\omega_{substrate} \quad (5)$$

$$\Delta\omega_{substrate} = \beta \int_{300\ K}^{T} (\alpha_{SiO_2}(T') - \alpha_{MoS_2}(T'))\, dT' \quad (6)$$

where $\beta$ is the biaxial strain coefficient, $\alpha_{SiO_2}(T')$ and $\alpha_{MoS_2}(T')$ are the thermal expansion coefficients of $SiO_2$ and $MoS_2$ at any given temperature $T'$. The biaxial strain coefficient $\beta$ can be obtained from uniaxial strain dependent studies using the relation: $\beta = -2\omega_0\gamma$ [61,62], where $\gamma$ is the Grüneisen parameter of the corresponding phonon. Using thermal expansion coefficients of $SiO_2$ [61] and monolayer $MoS_2$ [46], we have estimated the substrate effect on the $E_{2g}$ and $A_{1g}$ phonons ($F_2$ and $F_3$ modes) of monolayer $MoS_2$. The $\gamma$-values associated with the $F_2$ and $F_3$ modes were obtained from uniaxial strain measurements by Conley *et al.* [63] and Rice *et al.* [64], respectively. Subtracting the substrate effect from our data for 1L-$MoS_2$, we have obtained the actual temperature dependence of a free-standing $MoS_2$. As shown in Figure 8, the subtracted data show an appreciable change with respect to the substrate-supported data for the in-plane mode ($F_2$; $E_{2g}$) while the change is minimal in case of the out-of-plane mode ($F_3$; $A_{1g}$). The $\chi$-value so obtained for the $F_2$ mode ($E_{2g}$) matches well with the theoretical prediction for free-standing monolayer $MoS_2$ [35]. There are no reports of strain-dependent Raman studies on monolayer of $MoTe_2$ so far. Nonetheless, a recent report on strain-dependent Raman studies of multi-layer $MoTe_2$ reveal very weak dependence on strain [65]. Therefore, we may expect the substrate effect on the thermal studies performed on $MoTe_2$ to be minimal. The substrate correction for the $F_3$ mode ($E_{2g}$) has been computed for monolayer $MoTe_2$ as well and we observe a very negligible contribution in the thermal behaviour due to the substrate (See Supplemental note 5 of Supplemental material [43]).

**VIII. Polarization-angle dependence**

From the above discussions, we have a very clear understanding of the thermal behaviour of phonons both for the first-order and the second-order Raman processes. Finally, we will discuss the effect of polarization angle on the first-order and second-order Raman processes. Raman spectra were obtained in parallel (XX) and cross (XY) polarization configurations for the flakes of $MoS_2$ and $MoTe_2$ of various thicknesses (See Supplementary Figures S22-S24). The $MoS_2$ flakes were investigated with 532 nm (non-resonant) and 325 nm (resonant) excitation lasers, while the $MoTe_2$ flakes were probed using the 532 nm (resonant) excitation. It is clearly observed that all the out-of-plane modes and the second-order Resonant processes show very strong suppression in the cross-polarization configuration with respect to the parallel-polarization configuration for flakes of all thicknesses. However, the in-plane modes show very minimal dependence on the polarization direction. To further explore the polarization angle dependence, we have obtained Raman spectra of 2L-$MoS_2$ and 2L-$MoTe_2$ by rotating the polarization direction from 0°-360° and the corresponding polar plots are shown in Figures 9 and 10. In order to explain the behaviour exhibited by the different modes shown in Figures 9 and 10, we would need to refer to the Raman selection rules for polarization-angle dependent intensity as explained in Supplemental note S6 of the Supplemental material [43]. The intensity of a Raman mode may be calculated by the following expression:

$$I \propto \sum_j |e_s^t . R_j . e_i|^2 \quad (7)$$

where $e_i$ and $e_s$ are the unit vectors representing the polarization directions of the incident laser and the scattered Raman signal (the superscript t refers to the transpose of the $e_s$ matrix), and $R_j$ represents the Raman tensor corresponding to the given mode having a degeneracy j. Based on the degeneracy j, there may be multiple Raman tensors corresponding to a single mode. (e.g., the j values for $E$-type and $A$-type modes are 2 and 1, respectively). In such cases, the intensity of each of the tensors must be calculated using the expression above and the total intensity is given by the summation of the individual intensities. Using the appropriate Raman tensors for various modes, we obtain the following relations for the intensities of the $E_{1g}$, $E_{2g}$, and $A_{1g}$ modes:

$$I_{E_{1g}} = 0 \quad (8)$$

$$I_{E_{2g}} = |d|^2 \quad (9)$$

$$I_{A_{1g}} = |a \cos(\theta)|^2 \quad (10)$$

where $a$ and $d$ are constants and $\theta$ represents the polarization angle. As discussed in Section II, for 2L-MoS$_2$ flakes, the interlayer shearing mode (LSM) and the first-order in-plane mode F$_2$ correspond to the $E_{2g}$ symmetry and follow the equation (9). On the other hand, the interlayer breathing mode and the intralayer out-of-plane F$_3$ mode correspond to the $A_{1g}$ symmetry and follow the equation (10), exhibiting strong polarization angle dependence with double lobed shape in the polar plot (see Figure 9). The combination mode $\omega_3$ show similar polarization angle dependence as the $A_{1g}$-type modes, as also previously reported for MoS$_2$ [41] and WS$_2$ [66]. The non-resonant Raman measurements do not show the F$_1$ mode (possessing $E_{1g}$ symmetry), in accordance with the Raman selection rules described in equation (8). The resonant Raman measurements (of MoS$_2$) performed using the 325 nm laser excitation (Figure 9(b)) show the F$_1$ mode due to resonance-induced breaking of selection rules and it exhibits a similar behaviour as the F$_2$ and LSM modes possessing $E_{2g}$ symmetry. It may be noted here that Caramazza *et al.* [40] observed a similar behaviour for $E_{1g}$-type mode in edge-plane oriented bulk MoTe$_2$. However, the mode was not observed in the basal plane orientation in their study. Importantly, the in-plane modes, which should show no polarization dependence following equation (9) as discussed above, show observable polarization-angle dependence in the form of an anisotropy in the polar plots where the expected circular shape gets transformed into slightly elliptical. The anisotropic intensity for the in-plane modes can be attributed to layer-stacking misalignment, interlayer strain, and wavelength-dependent optical response of the beam-splitter [41,42]. In case of such an anisotropic behaviour, the polarization-angle (θ) dependent intensity (I) can be fitted with: [41,42]:

$$I_{E_{2g}} = |A \cos(\theta)|^2 + |B \sin(\theta)|^2 \quad (11)$$

The degree of anisotropy can be estimated by the parameter $\psi = A/B$, and is reported to show strong dependence on the flake-thickness, temperature, excitation wavelength, and resonance conditions [41,42]. We observe an increase in the anisotropy (perfect isotropy corresponds to $\psi$=1 and the value decreases with increase in the anisotropy) of the F$_2$ mode ($E_{2g}$) of MoS$_2$ in the polarization angle-dependent studies from $\psi$=0.98 for 532 nm excitation to $\psi$=0.87 for 325 nm laser excitation. It is also interesting to observe that the combinational modes also show very strong polarization-angle dependence. We observe that the combinational modes $\omega_8$ ($E_{1g} + E_{1g}$) and $\omega_{14}$ ($E_{2g} + A_{1g}$) show elliptical polar plots with $\psi$=0.47 and $\psi$=0.87, respectively, similar to the in-plane $E$-type modes. On the other hand, the $\omega_{13}$ ($2E_{2g}$) and $\omega_{15}$ ($2A_{1g}$) modes show double-lobe shape similar to $A_{1g}$ modes.

For MoTe$_2$, the resonance condition is already met with the 532 nm excitation laser and therefore, combinational and forbidden modes are observed in the polarization-angle dependent study. Similar to MoS$_2$, the in-plane modes F$_1$, F$_3$, and LSM exhibit minimal polarization angle dependence as compared to the out-of-plane modes. The in-plane modes show elliptical polar plots with varying degrees of anisotropy similar to the ones obtained for MoS$_2$ using resonant excitation of 325 nm. From the ratio of the coefficients obtained by fitting the data with equation 11, we observe varying degrees of anisotropy for the F$_1$, F$_3$, and LSM modes with $\psi$=0.80, 0.85, and 0.66, respectively. The F$_2$ and F$_4$ modes, both possessing $A_{1g}$ symmetry for 2L-MoTe$_2$ show double-lobe shaped polar plots, similar to MoS$_2$. The forbidden P$_3$ mode ($A_{1g}(M)$) also exhibits a double-lobed polar plot, as expected. The second-order mode $\omega_2$ ($E_{2g}^1(M) - TA(M)$) shows a behaviour similar to the in-plane modes, while all other second-order modes show double-lobed polar plots, similar to $A_{1g}$ phonons.

**Conclusion**

In summary, we have performed a systematic and comprehensive Raman study on MoS$_2$ and MoTe$_2$ which are potential candidates for the next generation nanoscale devices. The thermal behaviour of the different phonons has been studied as a function of flake-thickness. All the modes observed follow a linear temperature-dependence in the temperature range of 80-300 K. The first-order modes in MoS$_2$ show a stronger thickness sensitivity as compared to MoTe$_2$. The out-of-plane modes, i.e. $A_{1g}$ and $B_{2g}$ phonons in MoS$_2$ and MoTe$_2$, respectively, show stronger coupling with electrons as compared to the in-plane $E_{2g}$ phonons. The thickness dependence of the various first-order modes could be explained by consideration of electron-phonon interactions and phonon-phonon interactions. In order to elucidate the thermal behaviour of the second-order Raman processes, various excitation laser sources were used satisfying the required resonance conditions. The summation (combination) modes were observed to show a high value of temperature-coefficients which were close to the summation of the temperature coefficients of the respective individual (first-order) phonons. On the contrary, the difference modes show a very weak temperature susceptibility. The second-order Raman processes exhibit broad linewidths and minimal thickness sensitivity. The substrate effect on the thermal behaviour of phonons was also accounted for by estimating the mismatch in thermal expansion coefficients of the monolayer MoS$_2$ (or MoTe$_2$) flakes and the underlying SiO$_2$ substrate that results in a biaxial strain on the flake. The estimates based on our data suggest that the in-plane modes are more sensitive to the strain as compared to the out-of-plane modes which can affect the thermal properties of the 2D layers. We have also performed extensive polarization angle resolved Raman studies on MoS$_2$ and MoTe$_2$ to observe the effects of polarization angle on the first-order and the second-order modes. We have observed that the first-order out-of-plane modes show much stronger polarization angle dependence compared to the in-plane modes. The in-plane modes, which show almost no polarization angle dependence in the non-resonant Raman processes (with non-resonant laser excitation), show observable dependence when the resonance conditions are satisfied. The second-order combinational modes show very strong polarization angle dependence similar to the first-order out-of-plane modes. We believe that our comprehensive study provides an insight with a better understanding of the thermal properties of the phonons in semiconducting 2H-MoS$_2$ and 2H-MoTe$_2$, playing important role in their potential applications.

**Acknowledgments**

S. S. acknowledges the funding from DST/SERB (Grants No. ECR/2016/001376 and No. CRG/2019/002668) and Nanomission (Grant No. SR/NM/NS-84/2016(C)). The authors also acknowledge the Central Instrumental Facility at IISER Bhopal for the atomic force microscopy measurements.

**Table-1: The Raman modes in 2H-MoS$_2$ and their assignments**

| Designation | Phonon assignment | Frequency (cm$^{-1}$) | Observable using excitation | | |
|---|---|---|---|---|---|
| | | | 325 nm (~ 3.81 eV) (Resonant) | 532 nm (~ 2.33 eV) (Non-resonant) | 633 nm (~ 1.96 eV) (Resonant) |
| F$_1$ | $E_{1g}$ | 286 | ✓ | ✗ | ✗ |
| F$_2$ | $E_{2g}$ | 384 | ✓ | ✓ | ✓ |
| F$_3$ | $A_{1g}$ | 405 | ✓ | ✓ | ✓ |
| F$_4$ | $A_{1g}^2$ | 472 | ✓ | ✗ | ✗ |
| ω$_1$ | $A_{1g} - LA$ | 178 | ✗ | ✗ | ✓ |
| ω$_2$ | $c$ | 425 | ✗ | ✗ | ✓ |
| ω$_3$ | $2 LA$ | 457 | ✗ | ✓ | ✓ |
| ω$_4$ | $d$ | 467 | ✗ | ✗ | ✓ |
| ω$_5$ | $E_{2g} + TA$ | 526 | ✗ | ✗ | ✓ |
| ω$_6$ | $A_{1g} + TA$ | 570 | ✗ | ✗ | ✓ |
| ω$_7$ | $E_{2g} + LA$ | 601 | ✗ | ✗ | ✓ |
| ω$_8$ | $E_{1g} + E_{1g}$ | 630 | ✗ | ✗ | ✓ |
| ω$_9$ | $A_{1g} + LA$ | 640 | ✗ | ✗ | ✓ |
| ω$_{10}$ | $E_{1g} + 2 XA$ | 646 | ✗ | ✗ | ✓ |
| ω$_{11}$ | $2 E_{2g} (M_1)$ | 724 | ✗ | ✗ | ✓ |
| ω$_{12}$ | $2 E_{2g} (M_2)$ | 738 | ✗ | ✗ | ✓ |
| ω$_{13}$ | $2 E_{2g}$ | 756 | ✓ | ✗ | ✗ |
| ω$_{14}$ | $E_{2g} + A_{1g}$ | 780 | ✓ | ✗ | ✗ |
| ω$_{15}$ | $2 A_{1g}$ | 820 | ✓ | ✗ | ✗ |

**Table-2: Assignment of the Raman modes in 2H-MoTe$_2$**

| Designation | Phonon assignment | Frequency (cm$^{-1}$) |
|---|---|---|
| F$_1$ | $E_{1g}$ | 118 |
| F$_2$ | $A_{1g}$ | 171 |
| F$_3$ | $E_{2g}$ | 236 |
| F$_4$ | $B_{2g}$ | 290 |
| P$_1$ | TA (M) | 69 |
| P$_2$ | LA (M) | 101 |
| P$_3$ | LO, $A_{1g}$ (M) | 162 |
| P$_4$ | LO, $E_{2g}$ (M) | 247 |
| ω$_1$ | 2 TA(M) or $E^1_{2g}(M)$ - LA(M) | 138 |
| ω$_2$ | $E^1_{2g}(M)$ - TA(M) | 184 |
| ω$_3$ | $E_{1g}(M)$ + TA(M) or 2 LA(M) | 205 |
| ω$_4$ | $A_{1g}(M)$ + LA(M) | 264 |
| ω$_5$ | 2 $A_{1g}(M)$ or $E^1_{2g}(M)$ + TA(M) | 329 |
| ω$_6$ | $E^1_{2g}(M)$ + LA(M) | 347 |

**Table-3: Temperature coefficients of Raman modes in MoS$_2$ flakes of various thicknesses probed using 325 and 532 nm laser excitation sources**

(F: Forbidden)

| Designation | Temperature coefficient $\chi$ (cm$^{-1}$K$^{-1}$) | | | | |
|---|---|---|---|---|---|
| | 1L | 2L | 3L | 8L | Bulk |
| F$_1$ | F | -0.0161 | - | -0.0119 | -0.0145 |
| F$_2$ | -0.0161 | -0.0146 | -0.0135 | -0.0131 | -0.0118 |
| F$_3$ | -0.0242 | -0.0176 | -0.0187 | -0.0137 | -0.0115 |
| F$_4$ | -0.0199 | -0.0239 | - | -0.0217 | -0.0284 |
| $\omega_3$ | -0.0681 | -0.0264 | -0.0221 | -0.0156 | -0.0131 |
| $\omega_{13}$ | -0.0290 | -0.0367 | - | -0.0331 | -0.0378 |
| $\omega_{14}$ | -0.0380 | -0.0305 | - | -0.0342 | -0.0327 |
| $\omega_{17}$ | -0.0295 | -0.0357 | - | -0.0288 | -0.0289 |

**Table-4: Temperature coefficients of Raman modes in 8L-MoS$_2$ probed with 633 nm laser source**

| Designation | Frequency (cm$^{-1}$) | Temperature coefficient $\chi$ (cm$^{-1}$K$^{-1}$) |
|---|---|---|
| $\omega_2$ | 425 | -0.0399 |
| $\omega_3$ | 457 | -0.0284 |
| $\omega_4$ | 467 | -0.0224 |
| $\omega_5$ | 526 | -0.0113 |
| $\omega_6$ | 570 | -0.0145 |
| $\omega_7$ | 601 | -0.0170 |
| $\omega_8$ | 630 | -0.0393 |
| $\omega_9$ | 640 | -0.0125 |
| $\omega_{10}$ | 646 | -0.0090 |

**Table-5: Temperature coefficients for MoTe$_2$ flakes of different thicknesses (Laser excitation source: 532 nm)**

(F: Forbidden)

| Designation | Temperature coefficient $\chi$ (cm$^{-1}$K$^{-1}$) | | | |
|---|---|---|---|---|
| | 1L | 2L | 4L | Bulk |
| F$_1$ | F | -0.0057 | -0.0071 | F |
| F$_2$ | -0.0090 | -0.0088 | -0.0114 | -0.0054 |
| F$_3$ | -0.0124 | -0.0122 | -0.0110 | -0.0121 |
| F$_4$ | -0.0155 | -0.0164 | -0.0153 | -0.0156 |
| P$_1$ | -0.0055 | -0.0191 | - | - |
| P$_2$ | -0.0114 | -0.0141 | - | - |
| P$_3$ | - | -0.0040 | -0.0105 | -0.0165 |
| P$_4$ | -0.0126 | -0.0077 | -0.0142 | - |
| $\omega_1$ | -0.0127 | -0.0093 | -0.0143 | -0.0124 |
| $\omega_2$ | -0.0068 | -0.0085 | -0.0067 | -0.0046 |
| $\omega_3$ | -0.0110 | -0.0055 | -0.0087 | -0.0102 |
| $\omega_4$ | -0.0140 | -0.0141 | -0.0181 | - |
| $\omega_5$ | -0.0136 | -0.0146 | -0.0167 | -0.0128 |
| $\omega_6$ | -0.0169 | -0.0238 | -0.0153 | -0.0150 |

**Figures**

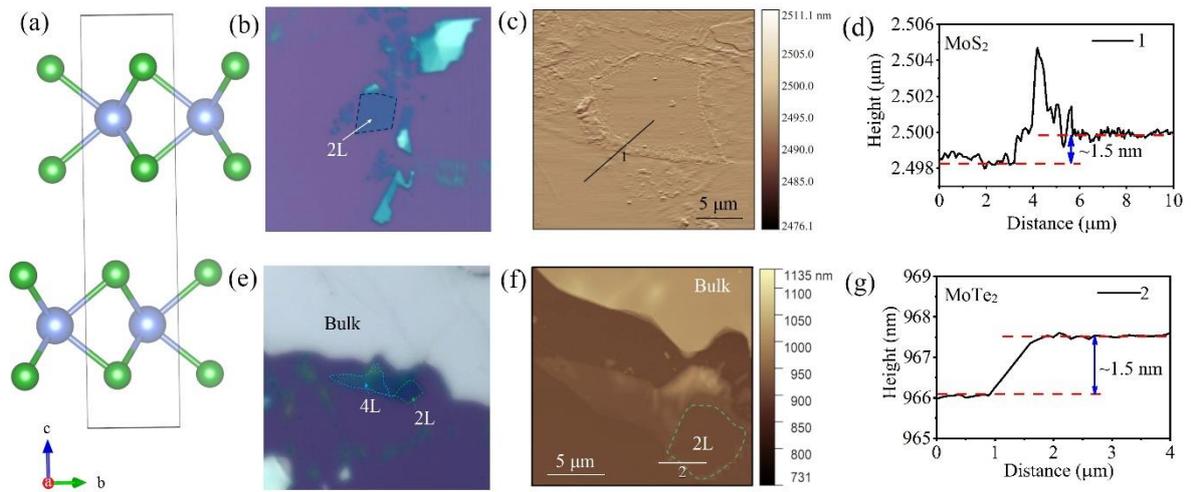

Figure 1. (Color online) (a) Atomic structure of 2H-MX$_2$ (M=Mo, X=S, Te). The big (blue) and small (green) spheres represent M and X atoms, respectively. (b,e) Optical microscope images of bilayer MoS$_2$ and MoTe$_2$, respectively. (c,f) The corresponding AFM images of bilayer MoS$_2$ and MoTe$_2$. (d,g) The height profiles across the lines marked 1 and 2 in (c) and (f) show the step heights of bilayer MoS$_2$ and MoTe$_2$, respectively.

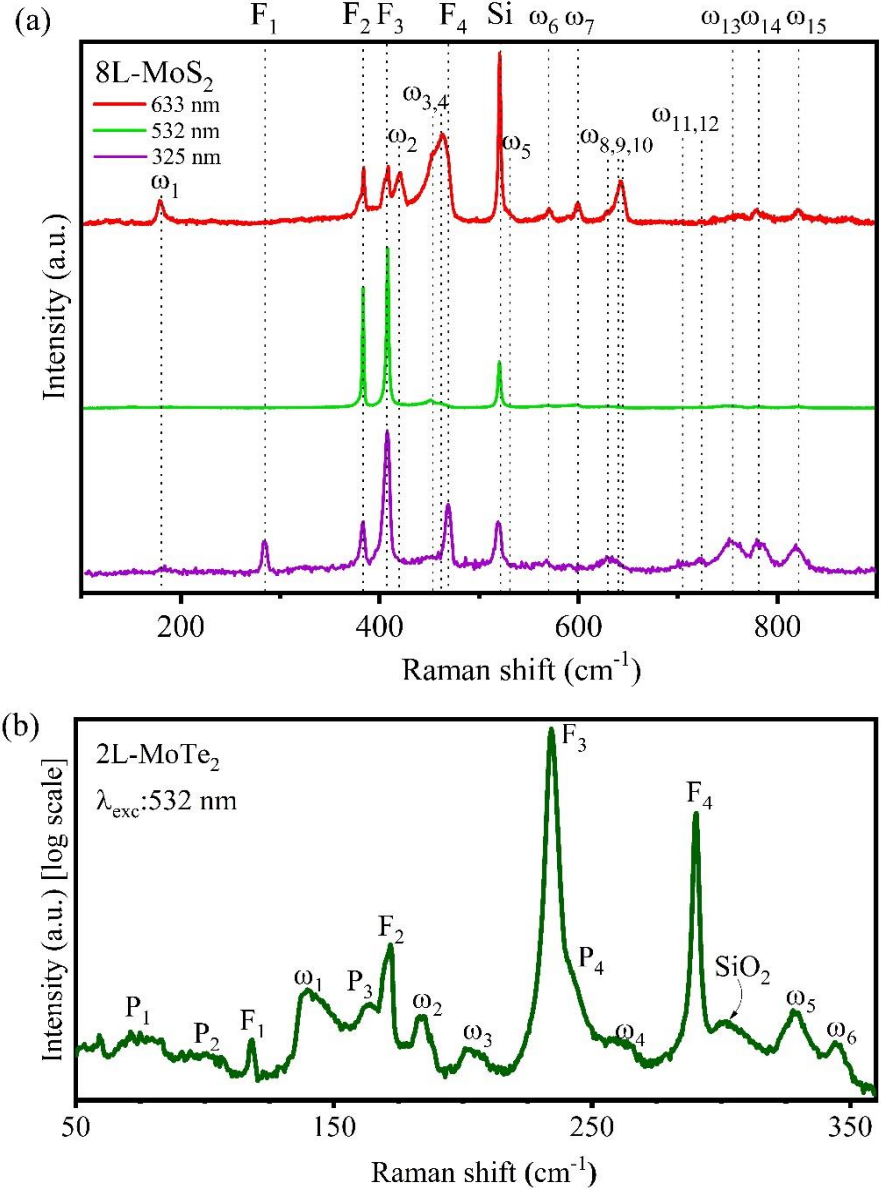

Figure 2. (Color online) (a) Room temperature Raman spectrum of 8L-$MoS_2$ obtained using three different laser excitation sources: 325 nm, 532 nm, and 633 nm. The modes are labelled as $F_1$ to $F_4$ (first-order modes) and $\omega_1$ to $\omega_{15}$ (combinational modes). The 532 nm excitation probes the first-order $F_2$ and $F_3$ modes. The first-order modes $F_1$ and $F_4$ and combinational modes at high frequency regions show resonance enhancement at 325 nm excitation, while other combinational modes ($\omega_1$ to $\omega_{10}$) show resonance enhancement at 633 nm excitation. (b) Room temperature Raman spectrum of bilayer $MoTe_2$ obtained using 532 nm excitation showing first-order ($F_1$ to $F_4$) phonons at the $\Gamma$-point, forbidden single-phonon processes ($P_1$ to $P_4$) away from the $\Gamma$-point, and the combinational modes ($\omega_1$ to $\omega_6$).

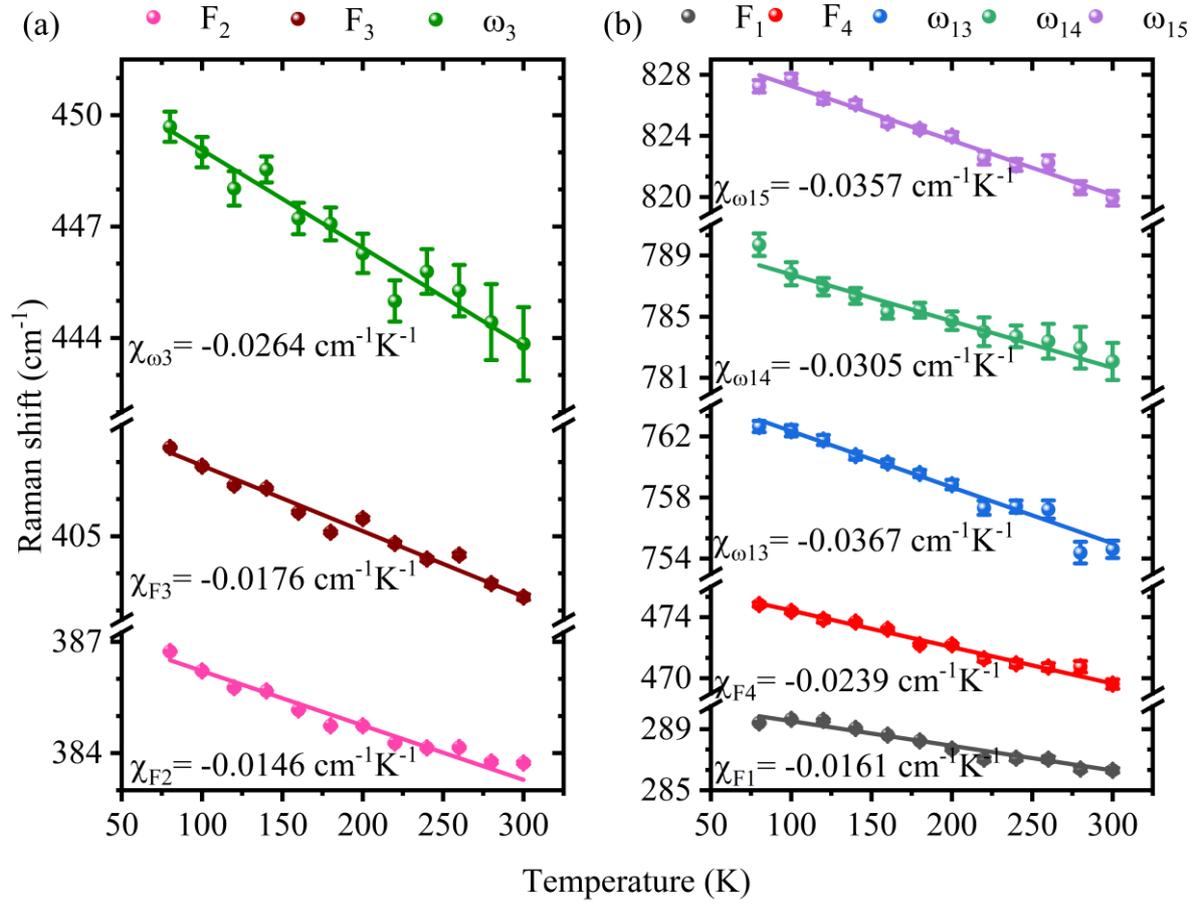

Figure 3. (Color online) The frequency vs temperature behaviours for the various phonons probed using (a) 532 nm and (b) 325 nm laser, for bilayer $MoS_2$. The corresponding temperature-coefficients ($\chi$) are also mentioned for each mode.

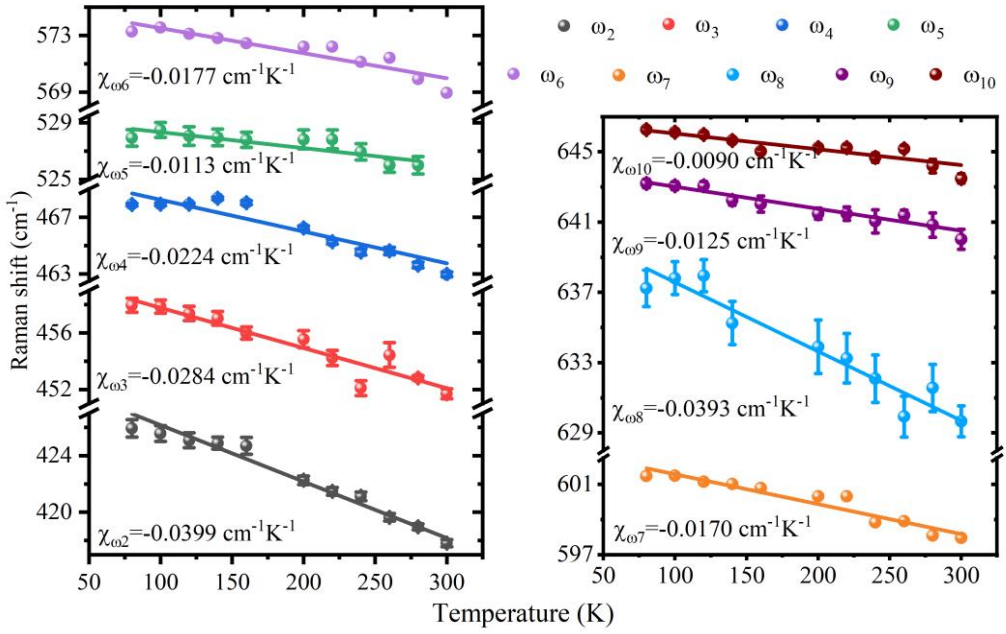

Figure 4. (Color online) The frequency vs temperature behaviours for the various phonons probed using 633 nm laser, for 8L-MoS$_2$. The corresponding temperature-coefficients ($\chi$) are also mentioned for each mode.

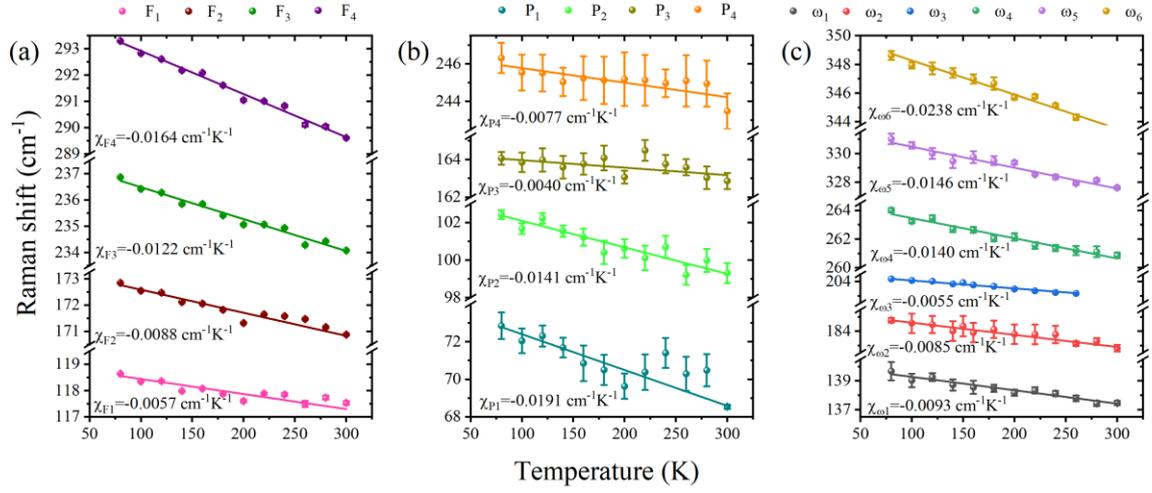

Figure 5. (Color online) The frequency vs temperature behaviours for (a) the first-order phonons ($F_1$ to $F_4$) at the $\Gamma$-point, (b) the forbidden single-phonon processes ($P_1$ to $P_4$) away from the $\Gamma$-point, and (c) the combinational modes ($\omega_1$ to $\omega_6$) due to double resonance processes at the $M$-point. The corresponding temperature-coefficients ($\chi$) are also mentioned for each mode.

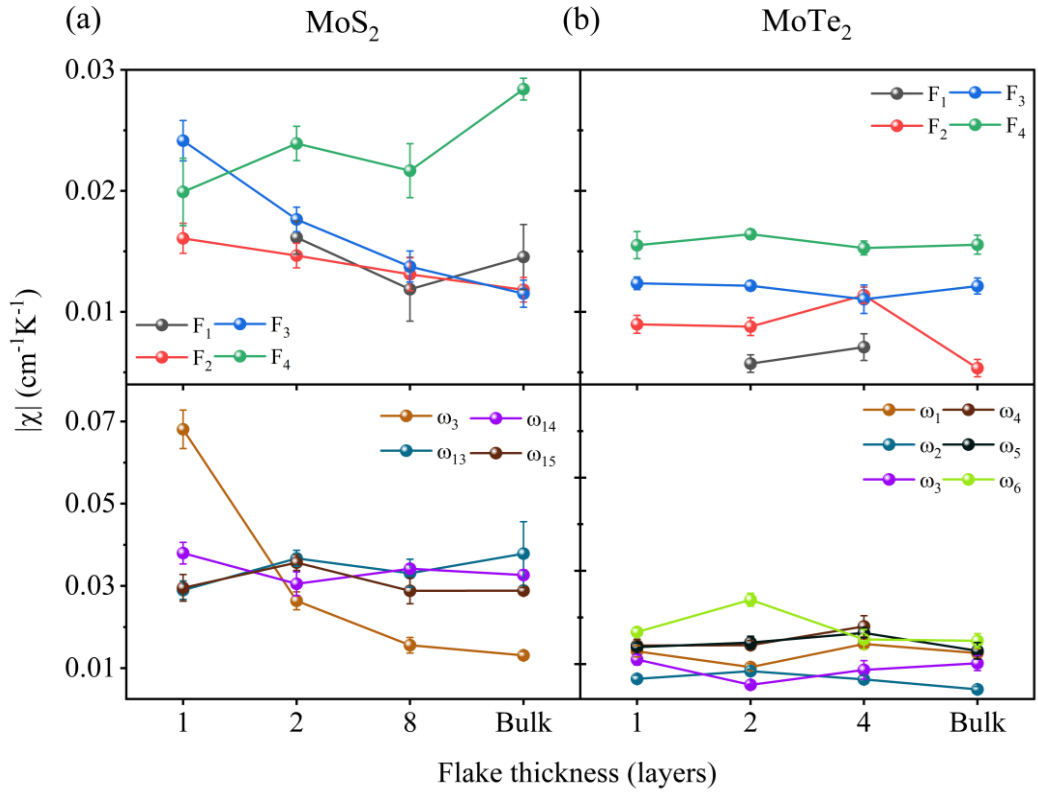

Figure 6. (Color online) The flake-thickness dependence of the temperature coefficients ($\chi$) obtained from the frequency vs temperature behaviours for the first-order and combinational modes of (a) $MoS_2$ and (b) $MoTe_2$. All the $\chi$ values obtained are negative and consistent with the redshift of phonons with increasing temperature. For the sake of comparison, the modulus values of the $\chi$'s are shown here.

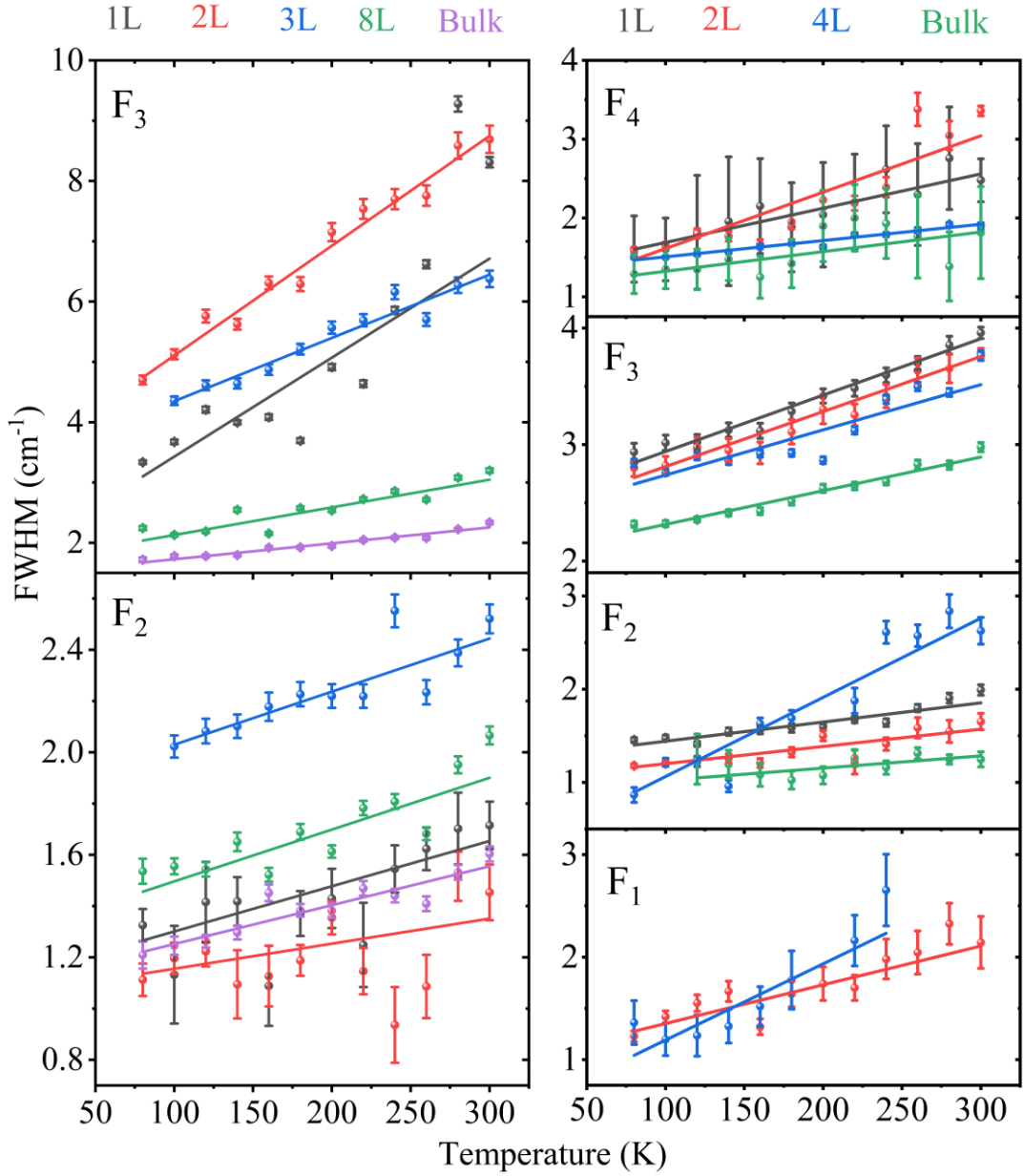

Figure 7. (Color online) The FWHMs (solid symbols) of the first-order Raman modes of (a) $MoS_2$ and (b) $MoTe_2$ probed using the 532 nm laser excitation. The temperature-susceptibility of the linewidths are represented by the slopes of the linear fits (solid lines) shown.

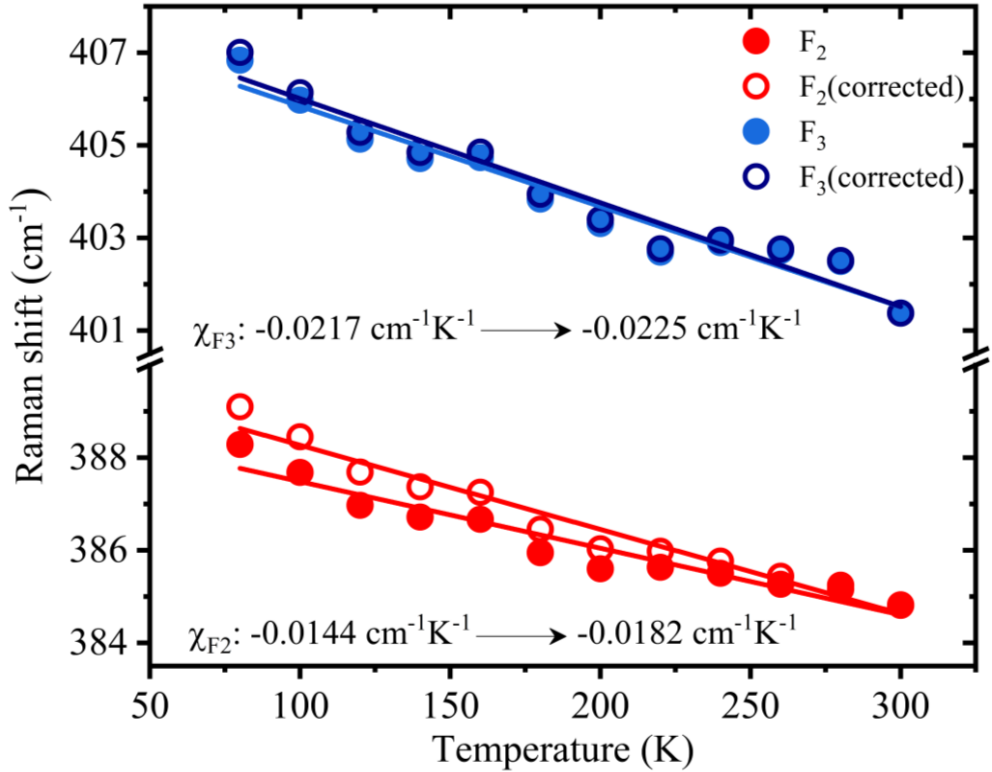

Figure 8. (Color online) The temperature dependence of the $E_{2g}$ ($F_2$) and $A_{1g}$ ($F_3$) phonons of 1L-MoS$_2$ when supported on SiO$_2$/Si substrate (solid circles). The phonon frequencies obtained after correcting for the substrate-effect using equations (5) and (6) described in text are shown with the open circles. Linear fittings of the data reveal a shift in the temperature coefficients, thereby revealing the original thermal effect on phonons. The substrate effect is more prominent on the in-plane ($F_2$: $E_{2g}$) mode as compared to the out-of-plane ($F_3$: $A_{1g}$) mode.

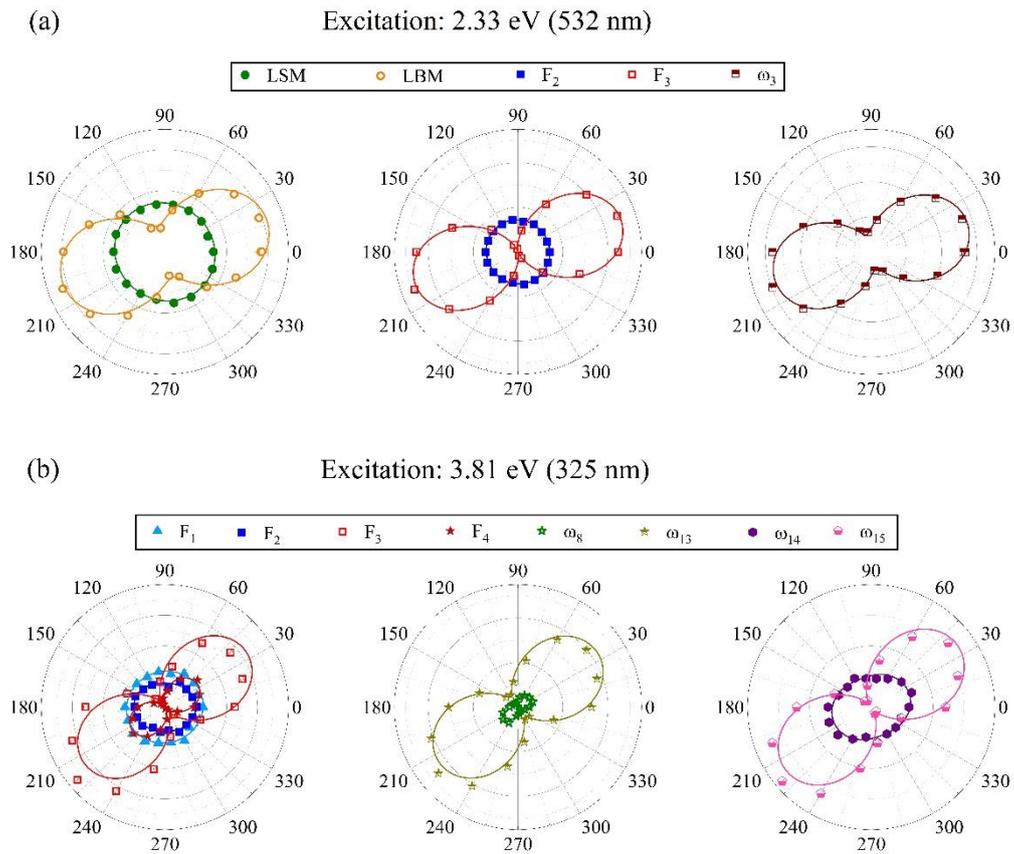

Figure 9. (Color online) The polar plots for the various modes of 2L-MoS$_2$ obtained using (a) 532 nm (non-resonant) and (b) 325 nm (resonant) excitation laser lines. The symbols represent the data points and the solid lines represent the suitable fittings using equations (10) and (11).

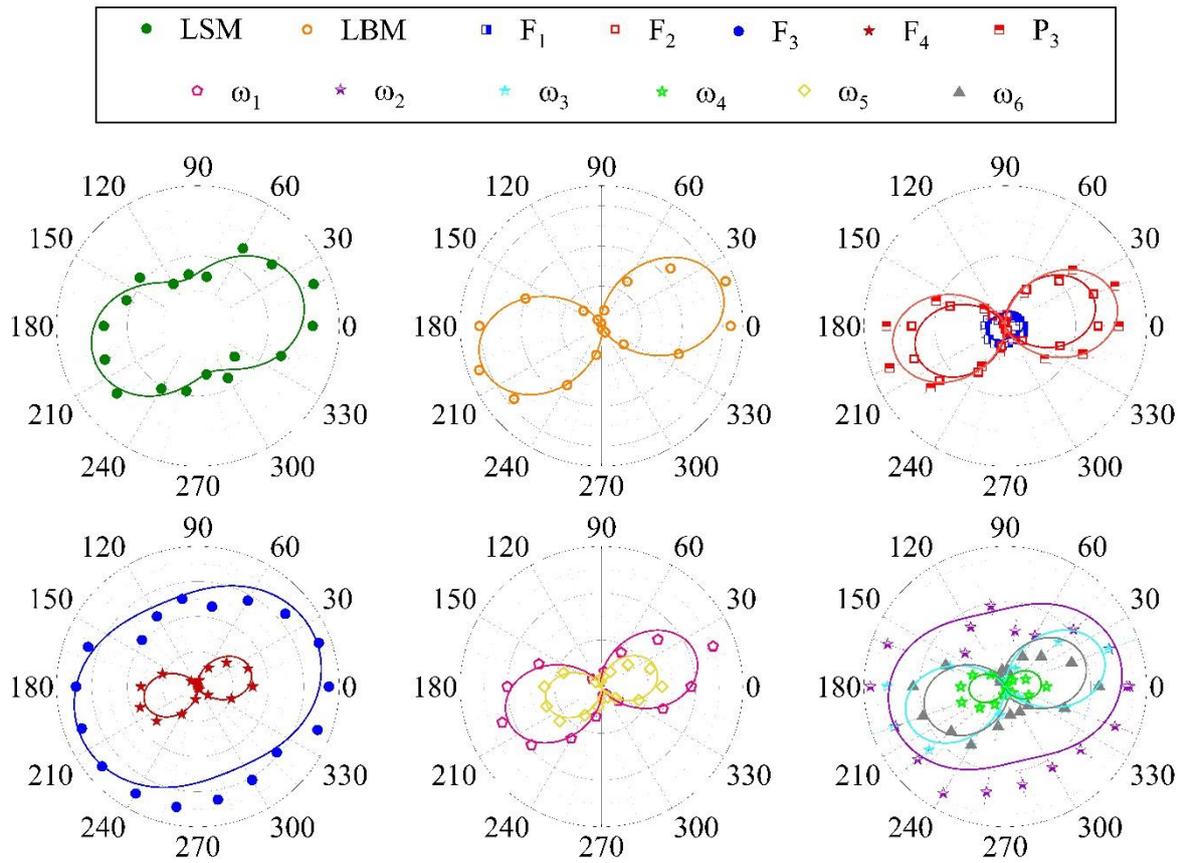

Figure 10. (Color online) The polar plots for the various modes of 2L-MoTe₂ obtained using 532 nm (resonant) excitation laser line. The symbols represent the data points and the solid lines represent the suitable fittings using equations (10) and (11).

**Electron-phonon and phonon-phonon anharmonic interactions in 2H-MoX$_2$ (X=S, Te): A comprehensive Resonant Raman study**

Suvodeep Paul, Saheb Karak, Annie Mathew, Ankita Ram, and Surajit Saha*

*Department of Physics, Indian Institute of Science Education and Research Bhopal, Bhopal 462066, India*

(*surajit@iiserb.ac.in)

**Supplemental material**

This Supplemental material contains additional discussions on:

- the thickness determination of the various investigated flakes of MoS$_2$ and MoTe$_2$ (with associated AFM and ultra-low frequency Raman data)
- resonance Raman enhancements of the various first-order and higher-order processes using variable excitation sources,
- temperature-dependent Raman (with comparison to previous reports)
- evidence of electron-phonon coupling
- the substrate-induced biaxial strain effect on thermal studies performed
- polarization-angle dependent Raman intensity and selection rules.

**Supplemental note 1**

*Thickness determination of MoS$_2$ and MoTe$_2$ flakes*

In the present work, the thermal behaviour of phonons was studied for various thicknesses of MoS$_2$ and MoTe$_2$. The thickness information of the investigated flakes of MoS$_2$ and MoTe$_2$ was obtained by various techniques. The optical microscope images of the flakes were obtained using a microscope fitted with 50× objective lens and the layer thicknesses could be determined accurately from the optical contrast. The thick bulk flakes appear white in colour, which changes from greenish to light blue as the flake-thicknesses are reduced to the atomically thin layers. The optical microscope images of the investigated flakes of MoS$_2$ and MoTe$_2$ of various thicknesses are shown in Figure S1(a) and Figure S2(a), respectively. Further confirmation of flake-thickness could be done from atomic force microscopy topographic images and the corresponding height profiles along the boundaries of the investigated flakes, as shown in Figures S1(c-f) and S2(c-e) for MoS$_2$ and MoTe$_2$, respectively. The height profiles for the monolayer flakes have not been shown as AFM measurements have been reported to overestimate the step height of monolayer flakes [1-4]. The final confirmation of the flake-thicknesses could be obtained from the room temperature Raman spectra which show thickness sensitive interlayer modes at the low frequency regions. There are two types of interlayer modes, *viz.* the interlayer shear mode and the interlayer breathing mode. The shear mode and breathing mode represent in-plane and out-of-plane interlayer vibrations for multilayer flakes and are absent for monolayer flakes [4-8]. Figures S1(b) and S2(f) show the low-frequency interlayer modes of MoS$_2$ and MoTe$_2$, respectively. As can be observed for both MoS$_2$ and MoTe$_2$ samples, the monolayer flakes show no interlayer modes. The bilayer flakes show an interlayer breathing and a shearing mode which flip their positions for thicker flakes. The breathing mode shows a redshift with increasing thickness and finally disappears for the bulk flakes. The shear mode, on the other hand, shows blueshift with increasing thickness till it saturates to the bulk shear mode frequency.

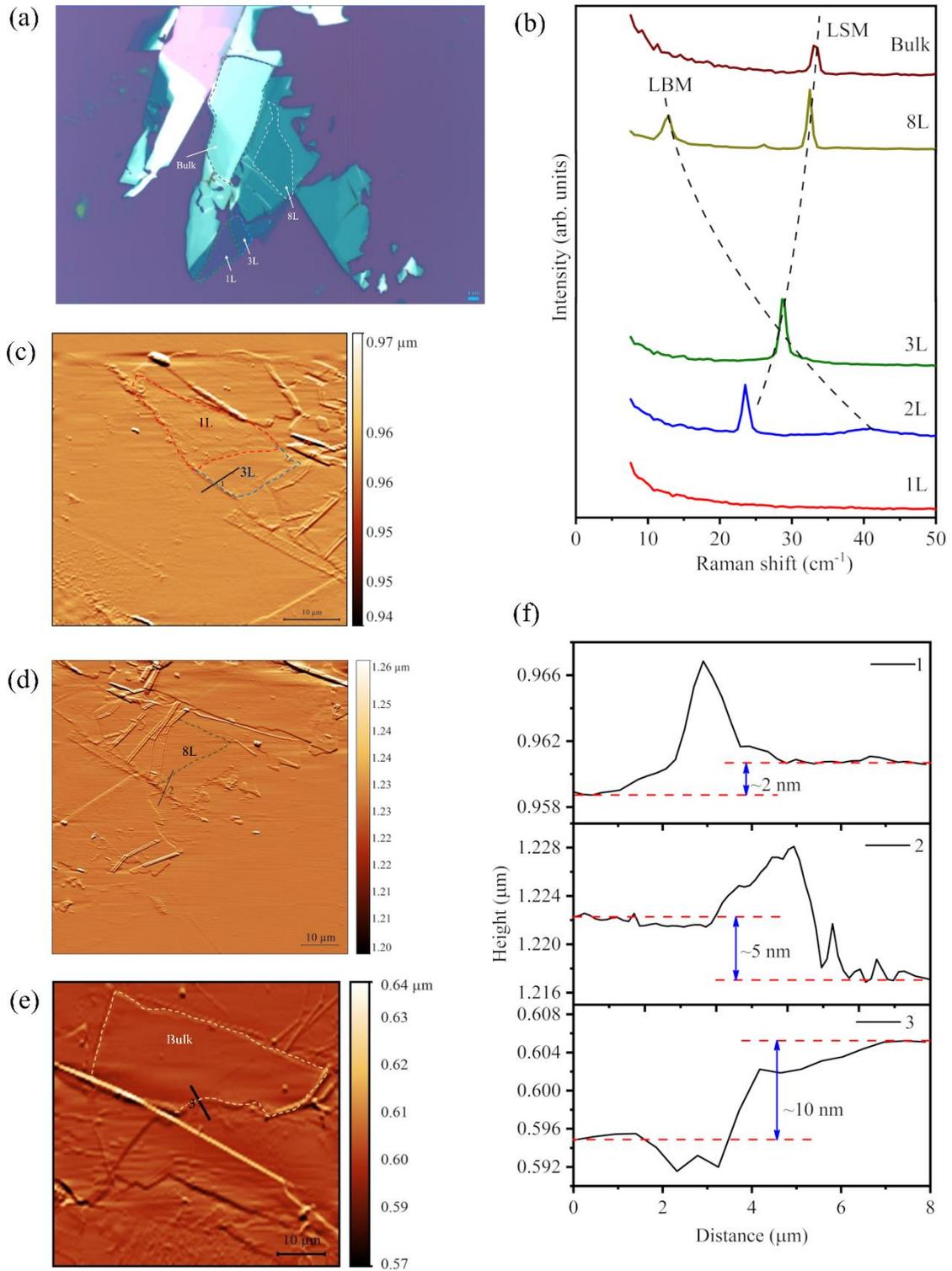

Figure S1 (a) Optical microscopic images of the various investigated flakes of $MoS_2$, (b) interlayer breathing (LBM) and shear (LSM) modes of $MoS_2$, confirming the flake-thicknesses, (c-e) AFM topographic images for different investigated layers, and (f) the corresponding height profiles revealing the flake-thicknesses.

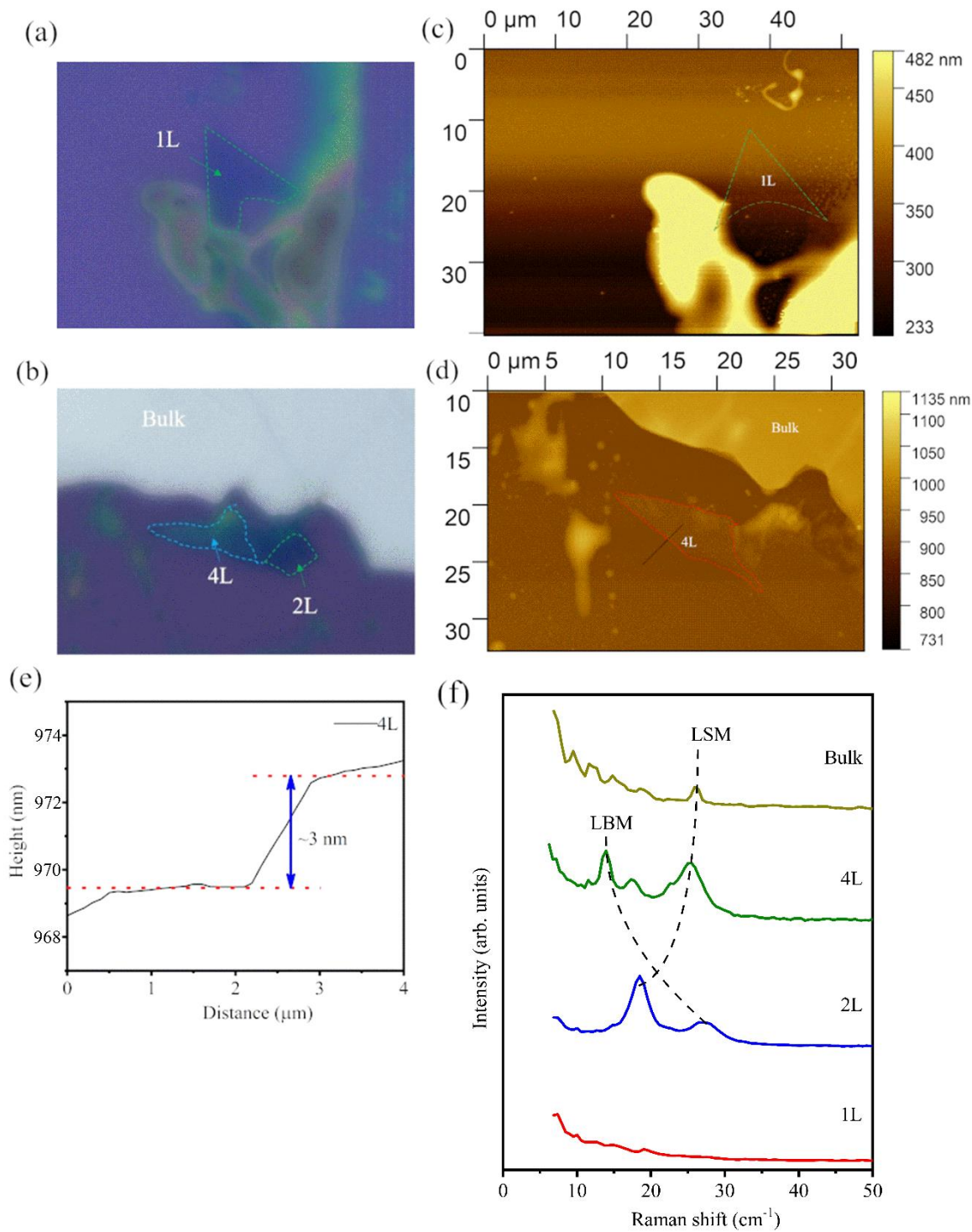

Figure S2 (a,b) Optical microscopic images of the various investigated flakes of MoTe$_2$, (c,d) AFM topographic images for different investigated layers, and (e) the corresponding height profile of the 4L flake revealing the flake-thickness, and (f) interlayer breathing (LBM) and shear (LSM) modes of MoS$_2$, confirming the flake-thicknesses. The 4L-flake also shows the secondary branches of the interlayer breathing and shear modes.

**Supplemental note 2**

*Excitation energy-dependent Raman spectra of MoS$_2$ and MoTe$_2$*

Figure S3 shows a stack of Raman spectra obtained at room temperature for various flake-thicknesses of MoS$_2$ using the 532 nm (2.33 eV) excitation laser source. We observe strong first-order modes F$_2$ and F$_3$ corresponding to the $E_{2g}$ and $A_{1g}$ phonons, respectively. The F$_2$ and F$_3$ modes show redshift and blueshift, respectively, with increasing flake-thickness. This behaviour has been explained by a competition between enhanced surface force constants due to stronger intralayer Mo-S interactions and a reduction of frictional force in thinner flakes of MoS$_2$ [9]. As a result, the frequency difference between the F$_2$ and F$_3$ phonons monotonically increases as a function of thickness with the minimum (~ 18 cm$^{-1}$) for monolayer and hence, can also be used for the confirmation of flake-thickness of MoS$_2$. The room temperature Raman spectra for various flake-thicknesses of MoS$_2$ obtained with 325 nm and 633 nm excitation lasers are shown in Figures S4 and S5, respectively. We observe that the forbidden first-order modes F$_1$ and F$_4$ and the high-frequency combinational modes $\omega_{11}$ to $\omega_{15}$ show strong resonance enhancement when excited with the 325 nm laser, which is consistent with previous reports [10, 11]. The modes F$_1$ and F$_4$ are forbidden by symmetry considerations for monolayer and bulk flakes of MoS$_2$. However, even for the multilayer flakes where these modes are allowed by symmetry rules, they appear when probed using the 325 nm laser only. This can be explained by the resonance of the excitation laser with the C exciton-type electronic transitions which have been reported to be spread over all the layers of MoS$_2$, unlike the A and B excitons which are confined to single layers [11]. On the other hand, there is no appreciable enhancement in the first-order modes F$_2$ and F$_3$. The combinational modes $\omega_1$ to $\omega_{10}$ involving phonons at the zone boundary show enhancement at 633 nm (1.96 eV) excitation due to resonance. The resonant combinational modes in the atomically thin layers of MoS$_2$ are barely observable because of the strong underlying photoluminescence feature, which is absent for the 8L flake. Therefore, 8L flake has been investigated for the thermal study of the combination modes. This may again be attributed to the confinement of the A excitons to single layers of MoS$_2$, resulting in a very strong photoluminescence feature for atomically thin flakes and being weak in thicker flakes. Further, the bandgap in atomically thin flakes is direct in nature whereas it becomes indirect in thicker flakes. Figure S6 shows the room temperature spectra of 8L-MoS$_2$ obtained using three different excitation laser sources. As can be observed from Figure S6 and from Figure 2(a) in the main text, the modes show variable enhancements when probed with various resonant (325 nm and 633 nm) and non-resonant (532 nm) laser excitation sources. Based on the figures S3-S6, the choice of suitable excitation wavelength for studying the thermal behaviours of various phonons in MoS$_2$ flakes was done. Quite obviously, the 532 nm excitation was used to probe the first-order F$_2$ and F$_3$ phonons. The 325 nm excitation was used to probe certain forbidden first-order modes (F$_1$ and F$_4$) and certain high frequency combinational modes that get enhanced because of the resonance effects mentioned above. The 633 nm laser was mainly useful for probing the combinational modes in the mid-frequency region (~200-750 cm$^{-1}$). Table 1 in the main text enlists the modes and the corresponding excitation wavelength(s) which causes their resonant enhancements.

Figures S7 and S8 show the room temperature spectra of MoTe$_2$ flakes of various thicknesses obtained using 532 nm and 633 nm laser excitations. Importantly, the electronic band structure of MoTe$_2$ has been reported to show two-dimensional Van Hove singularities (VHSs) at the M point corresponding to bandgaps of 2.30 eV and 2.07 eV. Therefore, we observe resonant enhancements for both 532 nm (2.33 eV) and 633 nm (1.96 eV) excitations. We observe comparable resonance enhancement for various investigated phonons when probed with either of the excitation energies. Therefore, we have performed the thermal studies on various flakes of MoTe$_2$ using the 532 nm excitation. Further, it was observed (see Figure S9) that the spectrum of bilayer MoTe$_2$ obtained using 325 nm (3.81 eV) excitation laser source show no resonant enhancement for most of the modes. Like MoS$_2$, the first-order modes F$_2$ and F$_3$ of symmetries $A_{1g}$ and $E_{2g}$, respectively, show an increasing frequency gap with increasing flake-thickness [12]. Additionally, the modes F$_1$ and F$_4$ (of symmetries $E_{1g}$ and $B_{2g}$, respectively) which are forbidden for monolayer and bulk MoTe$_2$ are observed for flakes of intermediate thicknesses. Due to strong resonance, some forbidden single phonon processes away from the $\Gamma$-point (P$_1$-P$_4$) could also be probed in monolayer flakes of MoTe$_2$, which showed very strong suppression in thicker flakes. The combination modes are summation modes or difference modes involving electron-phonon scattering from the inequivalent M points giving rise to double resonance processes. It is also important to note that while these combinational modes were previously observed for monolayer MoTe$_2$, we could probe them for thicker flakes as well.

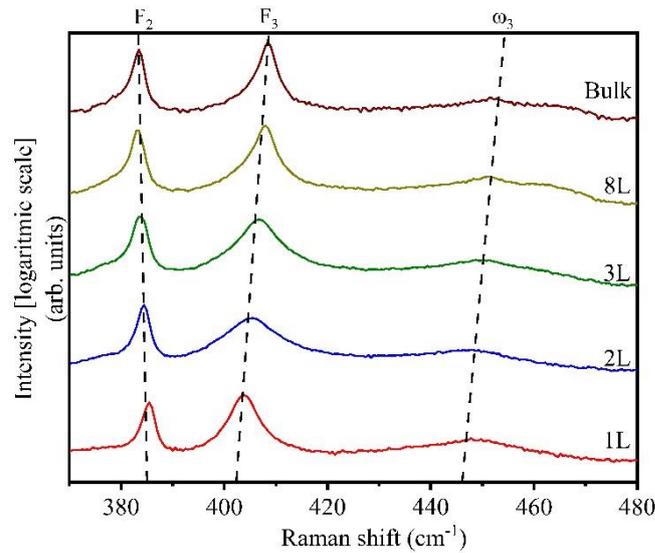

Figure S3. Stack of room temperature Raman spectra of MoS$_2$ layers of various thicknesses obtained using 532 nm (2.33 eV) excitation. The intensity axis is plotted in logarithmic scale for the sake of clarity so as to resolve the combinational mode $\omega_3$ which has a very low intensity relative to the first-order modes.

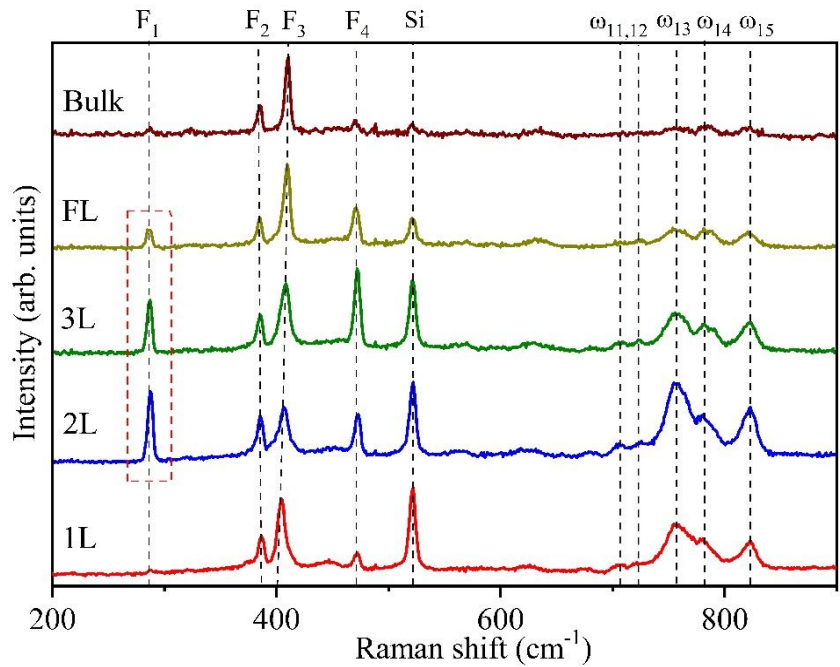

Figure S4. Stack of room temperature Raman spectra of MoS$_2$ layers of various thicknesses obtained using 325 nm (3.81 eV) excitation. The red dashed box shows the F$_1$ mode in the multilayer flakes, which is forbidden in the monolayer and bulk flakes. The F$_4$ mode, which is also forbidden in the monolayer and bulk flakes, is exhibited as a very weak mode in the monolayer and bulk flakes as well, which is due to the violation of the selection rules due to resonance with the C excitons.

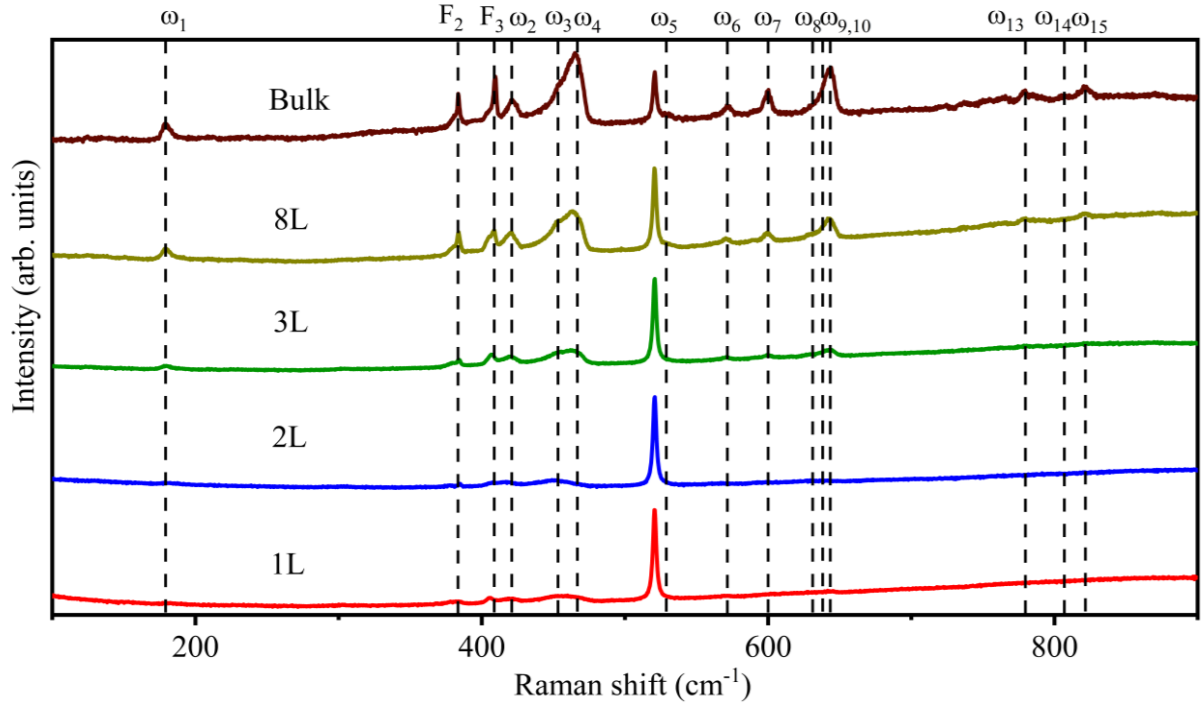

Figure S5. Stack of room temperature Raman spectra of $MoS_2$ layers of various thicknesses obtained using 633 nm (1.96 eV) excitation.

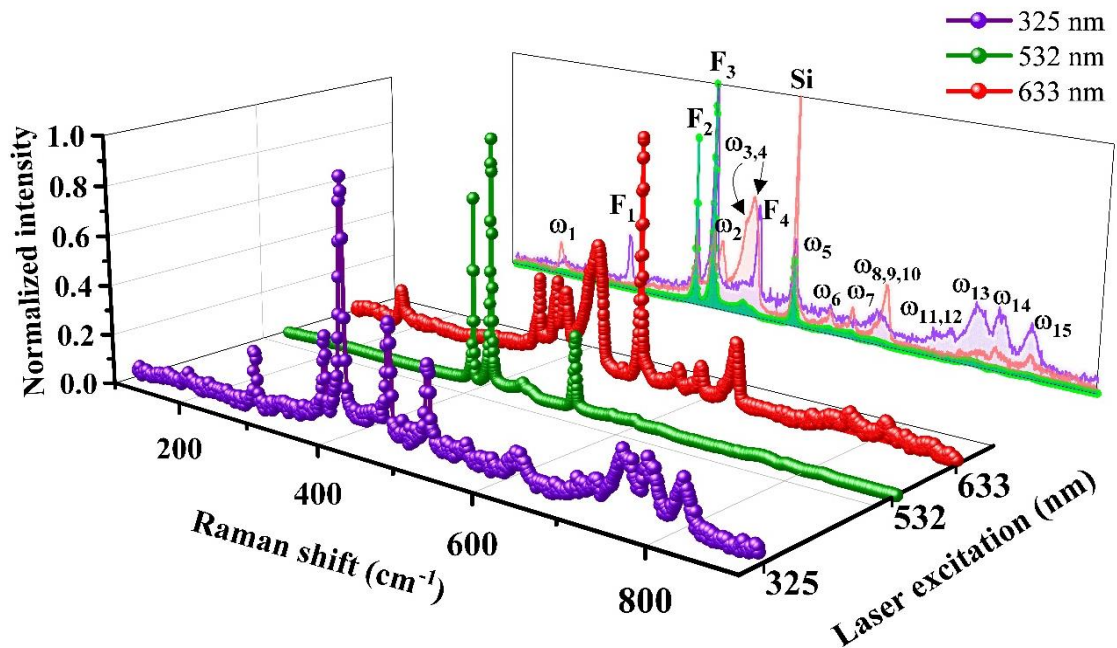

Figure S6. Comparison of room temperature Raman spectra of 8L-$MoS_2$ flake obtained using 325, 532, and 633 nm excitations. We observe variable intensity enhancements of various modes when probed with various resonant and non-resonant excitations.

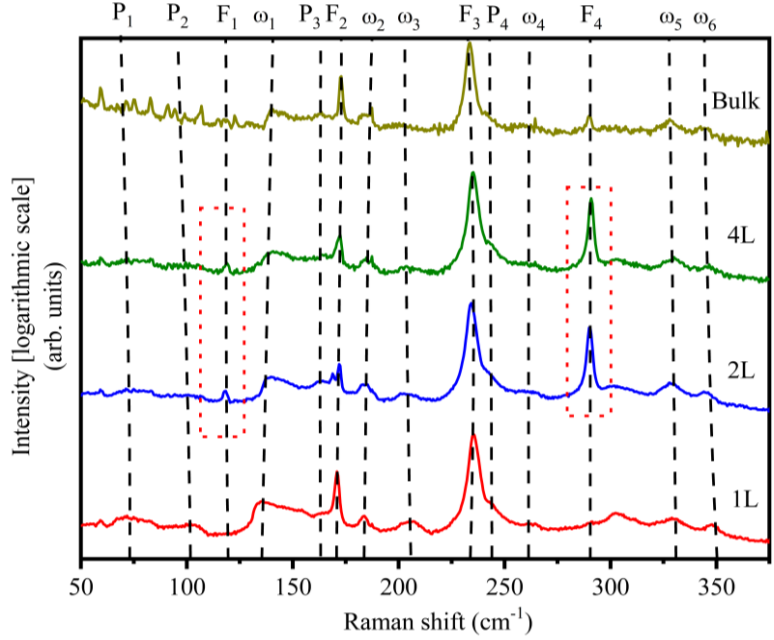

Figure S7. Stack of room temperature Raman spectra of MoTe$_2$ layers of various thicknesses obtained using 532 nm (2.33 eV) excitation. The red dashed boxes show the F$_1$ and F$_4$ modes in the multilayer flakes, which are forbidden in the monolayer and bulk flakes. The intensity axis is plotted in logarithmic scale to resolve the forbidden and the combinational modes which have very low intensities relative to the first-order modes.

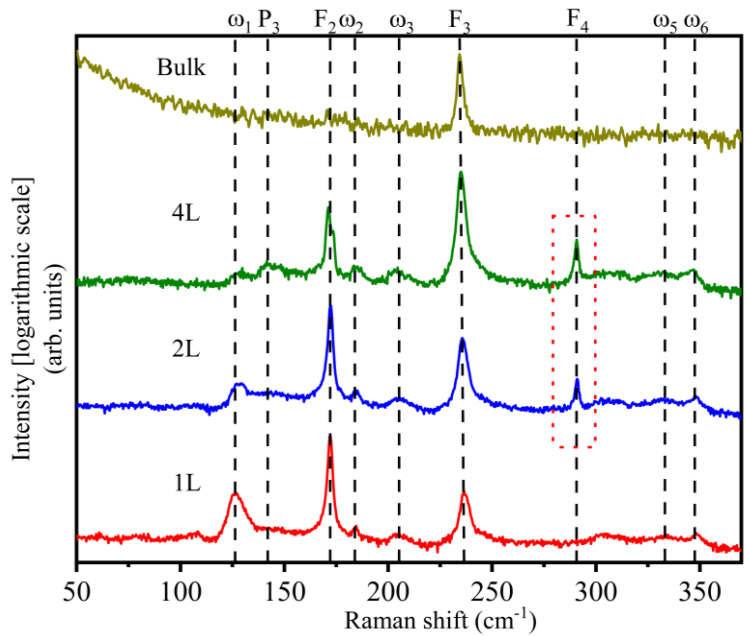

Figure S8. Stack of room temperature Raman spectra of MoTe$_2$ layers of various thicknesses obtained using 633 nm (1.96 eV) excitation. The Davydov splitting of the F$_2$ mode is clearly observed in the 4L-flake and is absent in all the other flakes. The red dashed box shows the F$_4$ mode in the multilayer flakes, which is forbidden in the monolayer and bulk flakes

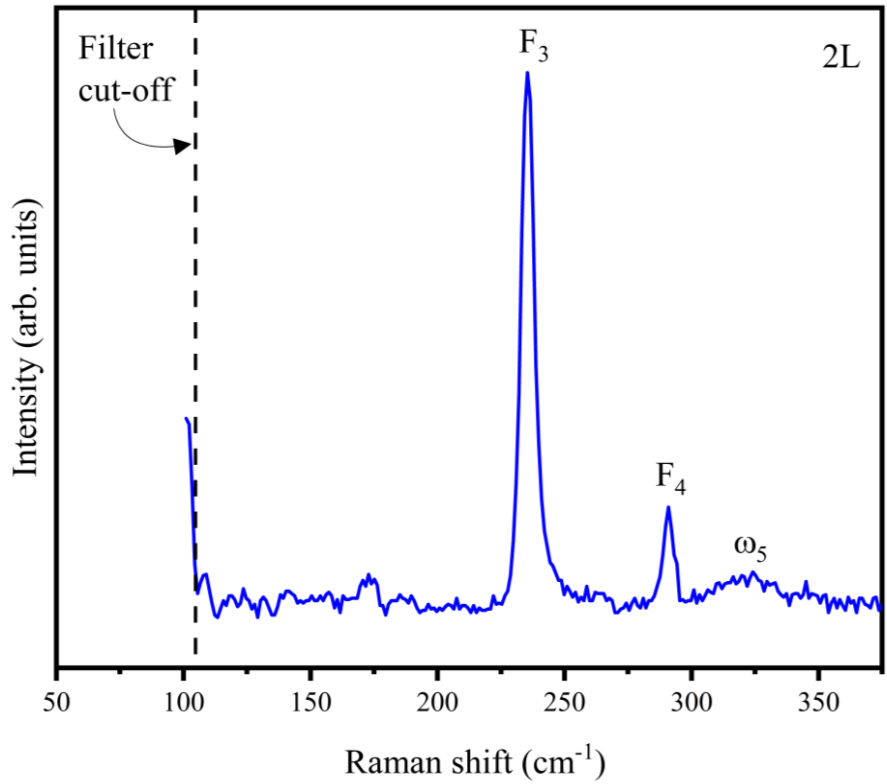

Figure S9. Room temperature Raman spectrum of bilayer MoTe$_2$ using 325 nm (3.81 eV) excitation.

**Supplemental note 3**

*Temperature-dependent Raman spectra of MoS₂ and MoTe₂*

The thermal behaviour of phonons of MoS$_2$ and MoTe$_2$ has been studied in great detail in the present work. Thermal studies have been performed for various thicknesses of MoS$_2$ and MoTe$_2$ with various excitations to quantitatively compare the behaviour of various phonons. Figures S10 and S11 show the temperature stacks for bilayer MoS$_2$ probed with 532 nm and 325 nm, respectively. Figure S12 shows the stack for 8L-MoS$_2$ (probed with 633 nm excitation laser), while Figure S13 shows the stack for bilayer MoTe$_2$ (probed with 532 nm excitation laser). It is observed that all the modes show a redshift in frequency as a function of temperature. As discussed in the main text, this behaviour is attributed to the volumetric expansion of the 2D flakes and the anharmonic phonon-phonon interactions. While there are various equations describing this behaviour depending on the channels of phonon decay and the order of the perturbation considered in the analysis, it has been reported that the temperature range from 80 K to 300 K effectively shows a linear temperature dependence of the frequency and linewidths of the phonons. We have fitted the frequency-temperature plots with the linear equation (3) of the main text and the corresponding linear temperature coefficients $\chi$ have been compared for various flake-thicknesses of MoS$_2$ and MoTe$_2$. Figure 3 and 5 of main text shows the frequency-temperature plots with corresponding $\chi$ values for bilayer MoS$_2$ and MoTe$_2$, respectively. Figures S14, S15, and S16 show the frequency-temperature plots of other investigated flakes of MoS$_2$ and MoTe$_2$, respectively. Temperature coefficients for certain first-order modes of MoS$_2$ and MoTe$_2$ were previously reported, while no data is available for the more complex processes (higher-order modes). The $\chi$ values obtained from our results match well with the previous reports for all the first-order modes. Table ST1 show a comparison of the $\chi$ values obtained from our results with earlier reports.

Finally, as has also been discussed in the main text, anharmonicity is also manifested in the phonons by a linewidth broadening as a function of temperature. The linewidth is interpreted as an inverse of the associated phonon lifetime and, hence, an increase in the linewidth corresponds to shortening of the phonon lifetimes. The linewidths associated with the modes also depend on the electron-phonon interactions. In our study, the linewidths of the various modes could be fitted with a linear equation as discussed in the main text. We have fitted the FWHMs (obtained from the Lorentzian fits of the various first-order modes of MoS$_2$ and MoTe$_2$ studied using the 532 nm excitation) using equation (4) of the main text, as also shown in Figure 7 of the main text. The corresponding slopes $\xi$ are shown in Figure S17. The generally observed trend for most of the phonons is a decrease in temperature sensitivity with increasing flake-thickness. This behaviour can be explained by considering the variation of the electron-phonon interactions as a function of flake-thickness. As discussed in main text, the electron-phonon interactions result in a decreasing trend of the phonon linewidth with increasing temperature. While all the modes show a positive slope ($\xi$) in the linewidth vs temperature plots, it is observed that the $\xi$-values decrease as a function of increasing flake-thickness. We can conclude that the addition of layers causes an increase in the electron population resulting in an increase in the electron-phonon interactions. Though the contribution from the anharmonic phonon-phonon interactions is still dominant irrespective of the layer thickness, the increase in electron-phonon interactions with increase in the flake-thickness results in an effective decrease in the $\xi$-values with increasing number of layers. We also observe in Figure S17 that the suppression in $\xi$-value in MoS$_2$ is stronger for the F$_3$ ($A_{1g}$) mode compared to the F$_2$ ($E_{2g}$) mode. This is because of a stronger coupling of the $A_{1g}$ phonon with the electrons as compared to the $E_{2g}$ phonons. In case of MoTe$_2$, we see an opposite trend in the flake-thickness dependence of the F$_1$ mode ($E_{1g}$ phonon), which has also been observed for the phonon frequency of this mode (Figure 6 of main text) and may be connected to the possible stronger resonance with the C excitons (which spread over all the layers and is not confined to a single layer) for thicker flakes discussed in main text. Finally, we also observe that the F$_2$ mode ($A_{1g}$ phonon) shows a sudden increase in the $\xi$ value for the 4-layer flake (also observed in the phonon frequency). In this regard, it may be noted that multilayer flakes of MoTe$_2$ have been reported to show the phenomenon of Davydov splitting of the out-of-plane $A_{1g}$ phonon [8]. While the phenomenon is not allowed in $A_{1g}$ phonons of monolayer and bilayer MoTe$_2$, the 4-layer flakes are supposed to show two closely spaced modes due to Davydov splitting [8]. This splitting is generally not very prominent with the 532nm excitation and are, therefore, studied using the 633 nm excitation. Consistent with previous reports [8], we observe a splitting of the $F_2$ mode in Figure S8 (room temperature spectra of different flakes of MoTe$_2$ obtained using 633 nm excitation) in case of the 4-layer thick flake. However, this splitting does not appear in the monolayer or bilayer flakes. On the other hand, the figure S7 shows no splitting in the $F_2$ mode of the 4-layer flake. We believe that the Davydov splitting of the $F_2$ mode could not be resolved by the 532 nm excitation due to instrumental resolution, but nevertheless, the phenomenon resulted in an effective broadening of the mode. This led to an overestimation of the FWHM for this mode.

**Table ST1:** Temperature coefficients of the first-order modes of MoS$_2$ and MoTe$_2$

| 2D-material | Thickness | Phonon | First-order temperature coefficient, $\chi$ (cm$^{-1}$K$^{-1}$) | Reference |
|---|---|---|---|---|
| MoS$_2$ | 1L | $E_{2g}$ | - 0.0179 | [13] |
| | | $A_{1g}$ | - 0.0143 | |
| | 2L | $E_{2g}$ | -0.0136 | |
| | | $A_{1g}$ | -0.0188 | |
| | 3L | $E_{2g}$ | -0.0126 | |
| | | $A_{1g}$ | -0.0180 | |
| | 4L | $E_{2g}$ | -0.0145 | |
| | | $A_{1g}$ | -0.0178 | |
| | 5L | $E_{2g}$ | -0.0146 | |
| | | $A_{1g}$ | -0.0151 | |
| | 6L | $E_{2g}$ | -0.0137 | |
| | | $A_{1g}$ | -0.0132 | |
| | 1L | $E_{2g}$ | -0.0130 | [14] |
| | | $A_{1g}$ | -0.0160 | |
| | Bulk | $E_{2g}$ | -0.0150 | |
| | | $A_{1g}$ | -0.0130 | |
| | 1L | $E_{2g}$ | -0.0110 | [15] |
| | | $A_{1g}$ | -0.0130 | |
| | 1L | $E_{2g}$ | -0.0124 | [16] |
| | | $A_{1g}$ | -0.0143 | |
| | 1L | $E_{2g}$ | -0.0175 | [17] |
| | | $A_{1g}$ | -0.0177 | |
| MoTe$_2$ | 1L | $E_{2g}$ | -0.0130 | [18] |
| | | $A_{1g}$ | -0.0093 | |
| | 12L | $E_{2g}$ | -0.0164 | |
| | | $A_{1g}$ | -0.0092 | |
| | | $B_{2g}$ | -0.0150 | |
| | Bulk | $E_{2g}$ | -0.0140 | |
| | | $A_{1g}$ | -0.0080 | |
| | 1L | $E_{2g}$ | -0.0119 | [19] |
| | 2L | $E_{2g}$ | -0.0116 | |
| | | $B_{2g}$ | -0.0181 | |
| | 3L | $E_{2g}$ | -0.0113 | |
| | | $B_{2g}$ | -0.0141 | |
| MoS$_2$ | 1L | $E_{2g}$ | -0.0161 | This work |
| | | $A_{1g}$ | -0.0242 | |
| | 2L | $E_{2g}$ | -0.0146 | |
| | | $A_{1g}$ | -0.0176 | |

| | 3L | $E_{2g}$ | -0.0135 | |
| | | $A_{1g}$ | -0.0187 | |
| | 8L | $E_{2g}$ | -0.0131 | |
| | | $A_{1g}$ | -0.0137 | |
| | Bulk | $E_{2g}$ | -0.0118 | |
| | | $A_{1g}$ | -0.0115 | |
| MoTe$_2$ | 1L | $E_{2g}$ | -0.0090 | This work |
| | | $A_{1g}$ | -0.0124 | |
| | | $B_{2g}$ | -0.0155 | |
| | 2L | $E_{1g}$ | -0.0057 | |
| | | $E_{2g}$ | -0.0088 | |
| | | $A_{1g}$ | -0.0122 | |
| | | $B_{2g}$ | -0.0164 | |
| | 4L | $E_{1g}$ | -0.0071 | |
| | | $E_{2g}$ | -0.0114 | |
| | | $A_{1g}$ | -0.0110 | |
| | | $B_{2g}$ | -0.0153 | |
| | Bulk | $E_{2g}$ | -0.0054 | |
| | | $A_{1g}$ | -0.0121 | |
| | | $B_{2g}$ | -0.0156 | |

**Table ST2:** Temperature coefficients of the linewidths of the first-order modes of MoS$_2$ and MoTe$_2$

| Material | Phonon designation | Temperature coefficients $\xi$ of the linewidths in flakes of thickness (cm$^{-1}$K$^{-1}$) | | | | | |
|---|---|---|---|---|---|---|---|
| | | 1L | 2L | 3L | 4L | 8L | Bulk |
| 2H-MoS$_2$ | $F_2$ | 0.0018 | 0.0010 | 0.0021 | - | 0.0020 | 0.0015 |
| | $F_3$ | 0.0164 | 0.0182 | 0.0105 | - | 0.0046 | 0.0026 |
| 2H-MoTe$_2$ | $F_1$ | - | 0.0038 | - | 0.0074 | - | |
| | $F_2$ | 0.0021 | 0.0018 | - | 0.0085 | - | 0.0013 |
| | $F_3$ | 0.0048 | 0.0047 | - | 0.0039 | - | 0.0029 |
| | $F_4$ | 0.0044 | 0.0072 | - | 0.0021 | - | 0.0025 |

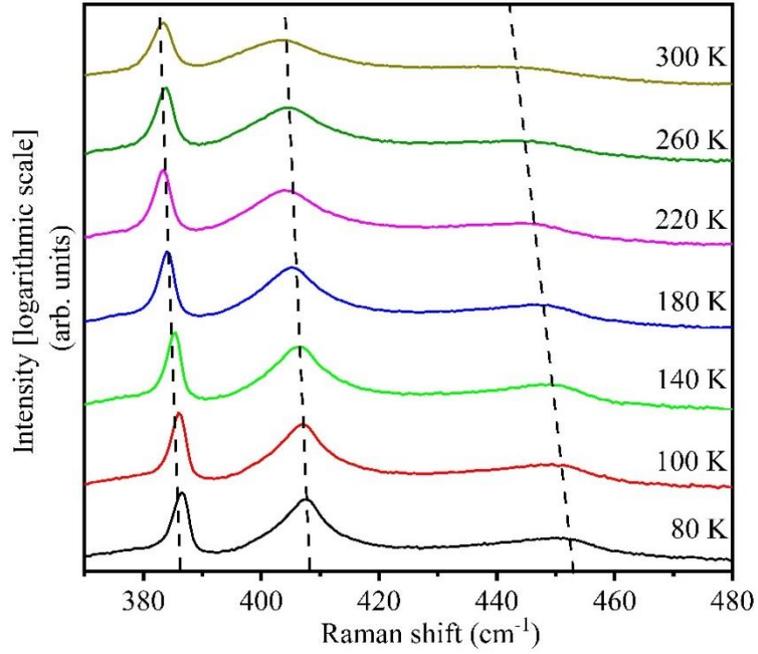

Figure S10. Stack of Raman spectra of 2L-MoS$_2$ obtained using 532 nm (2.33 eV) excitation showing redshift of phonons with increasing temperature. The intensity axis is plotted in logarithmic scale to resolve the combinational mode $\omega_3$ which has a very low intensity relative to the first-order modes.

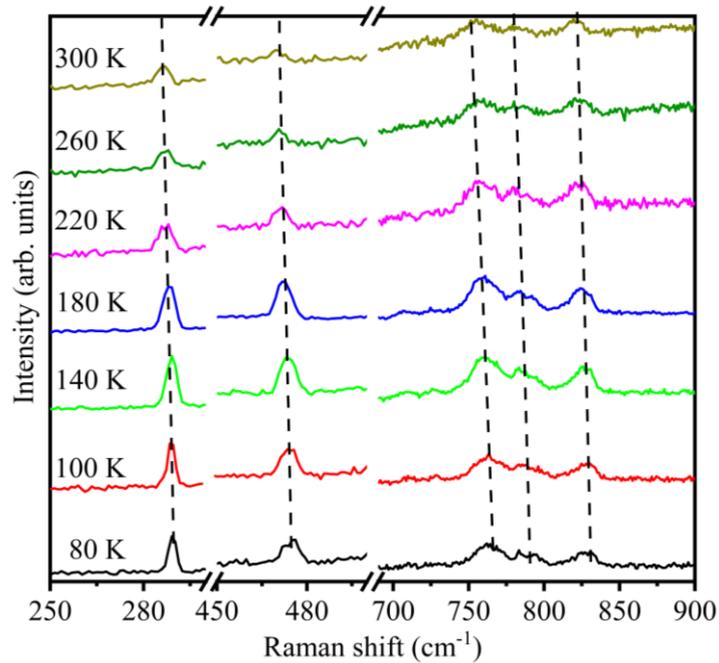

Figure S11. Stack of Raman spectra of 2L-MoS$_2$ obtained using 325 nm (3.81 eV) excitation showing redshift of phonons with increasing temperature.

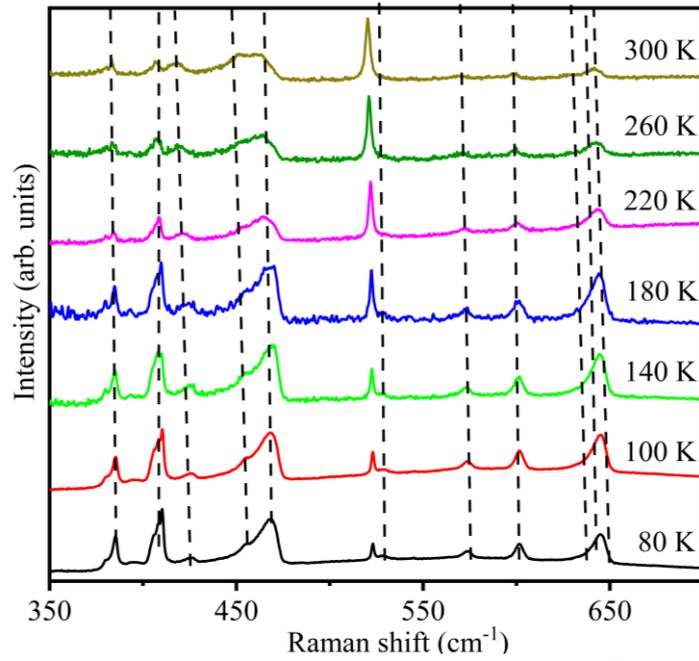

Figure S12. Stack of Raman spectra of 8L-MoS$_2$ obtained using 633 nm (1.96 eV) excitation showing redshift of phonons with increasing temperature.

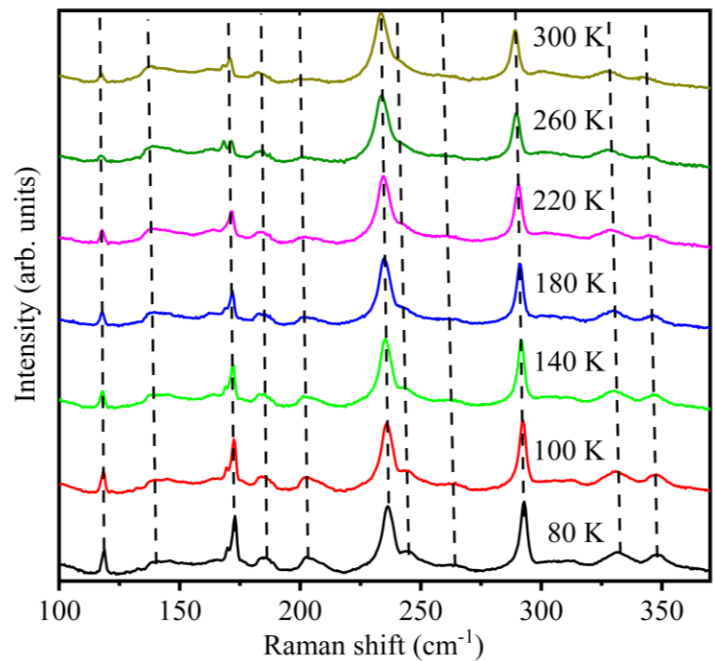

Figure S13. Stack of Raman spectra of 2L-MoTe$_2$ obtained using 532 nm (2.33 eV) excitation showing redshift of phonons with increasing temperature.

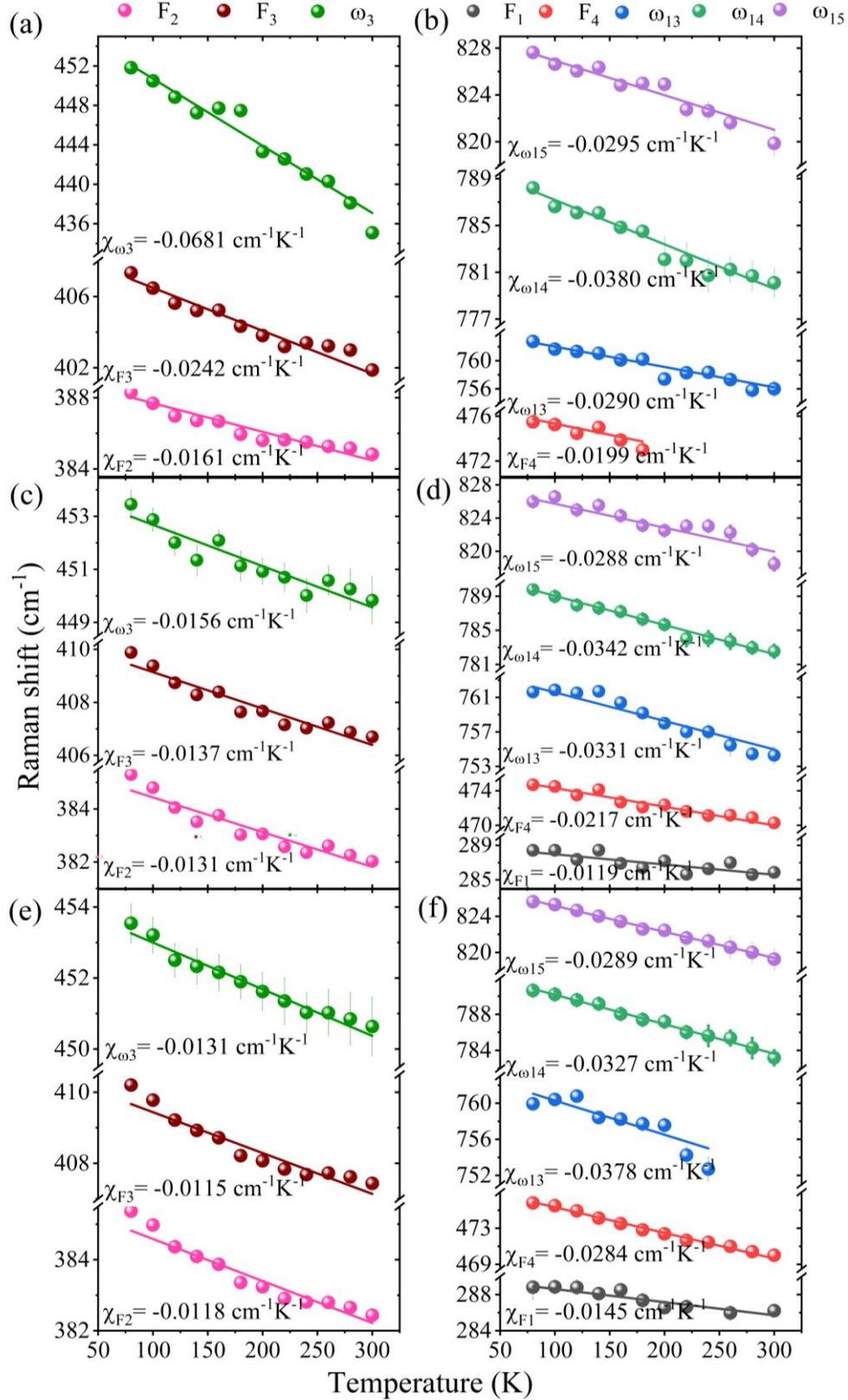

Figure S14. Frequency-temperature plots with linear fittings for (a,b) monolayer, (c,d) 8-layer, and (e,f) bulk MoS$_2$, respectively. The first column (a, c, and e) represents the modes probed using the 325 nm laser excitation, while the second column (b, d, and f) represents the modes probed using 532 nm laser excitation.

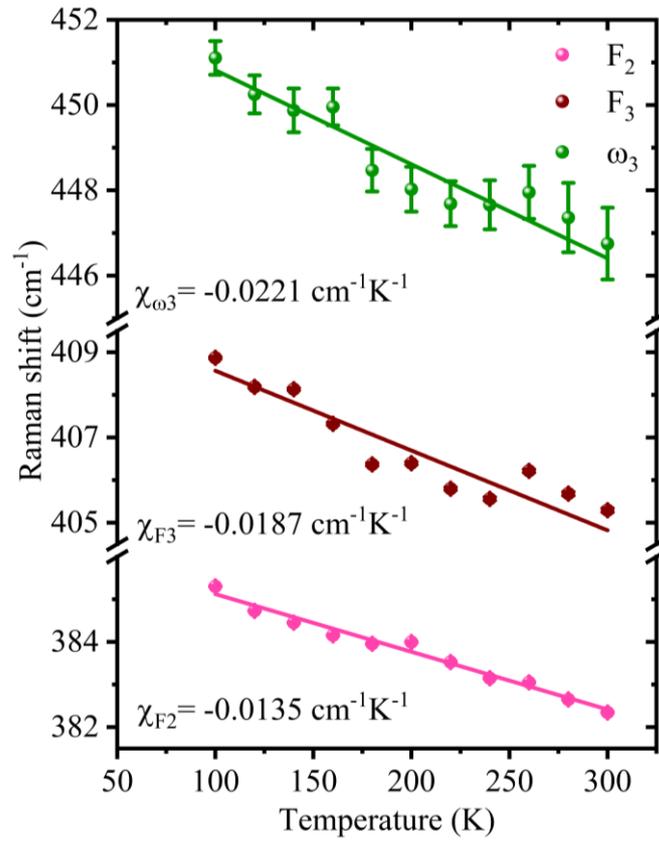

Figure S15. Frequency-temperature plots with linear fittings for trilayer MoS$_2$ using 532 nm excitation.

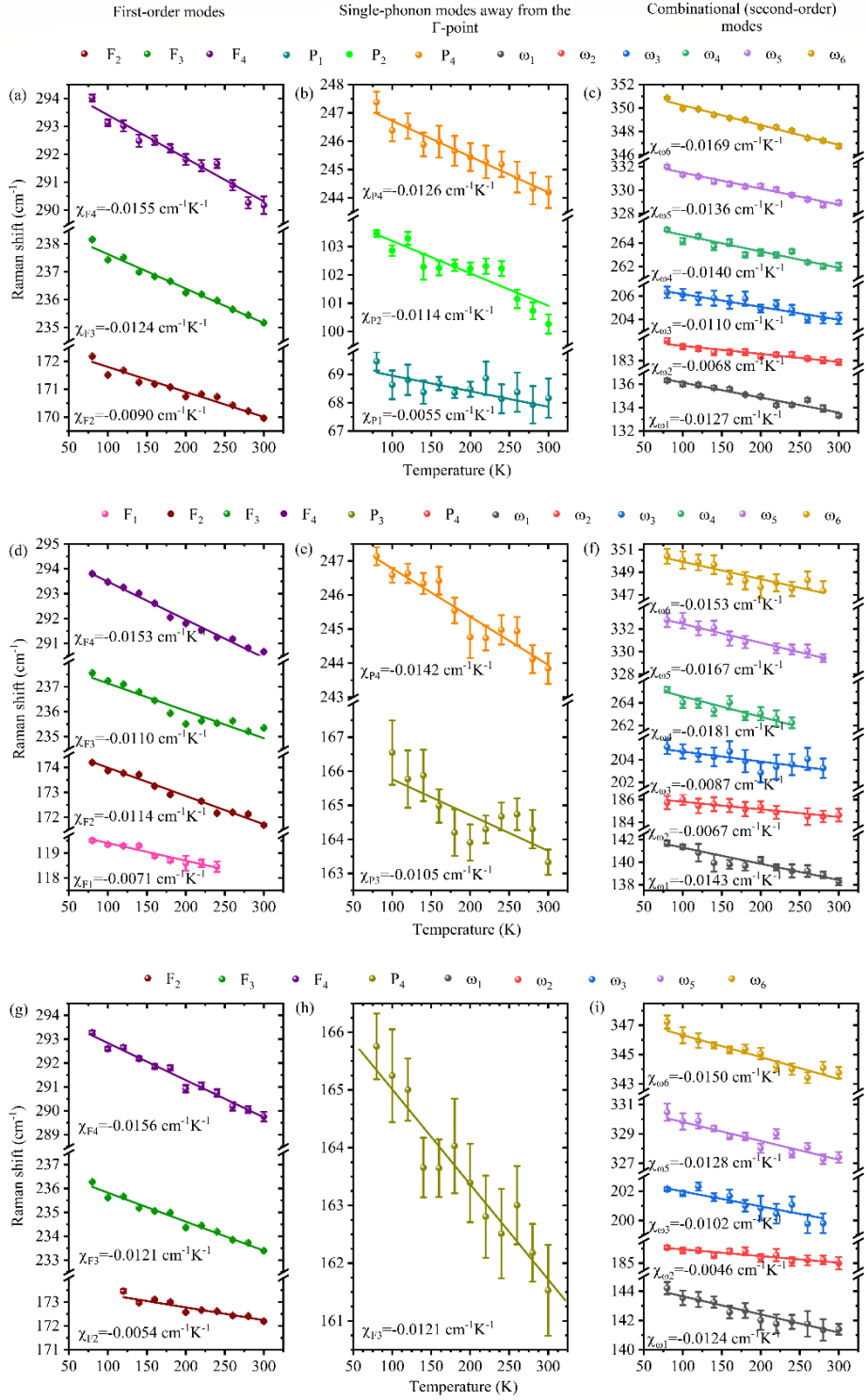

Figure S16. Frequency-temperature plots with linear fittings for (a-c) monolayer, (d-f) 4-layer, and (g-i) bulk MoTe$_2$, respectively. The first column (a, d, and g) shows the plots of the first-order modes, while the second column (b, c, and h) shows the forbidden single-phonon modes away from the Γ-point, and the third column (c, f, and i) represents the combinational modes.

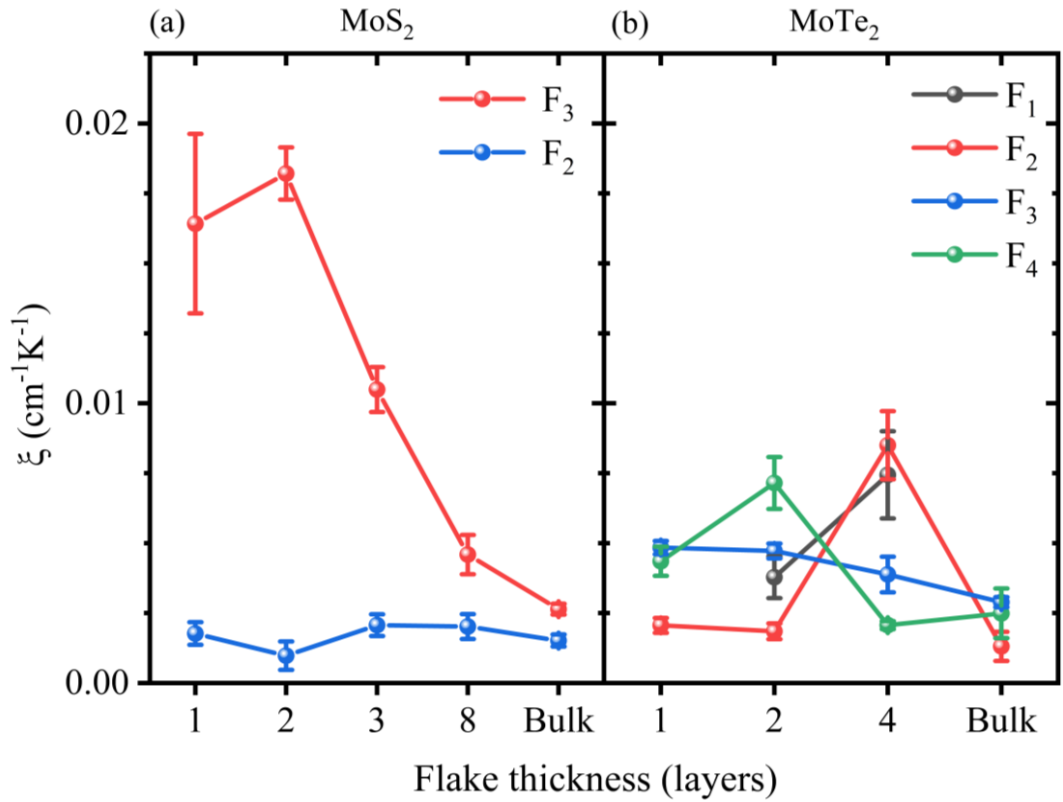

Figure S17. The flake-thickness-dependence of first-order temperature coefficients ($\xi$) obtained from FWHM vs temperature plots for (a) $MoS_2$ and (b) $MoTe_2$ using 532 nm excitation. We observe an overall suppression of the $\xi$-values with increasing layer-thickness for $MoS_2$ and $MoTe_2$, which can be attributed to an increase in the electron-phonon interactions. The effect is stronger in $MoS_2$ as compared to $MoTe_2$.

**Supplemental note 4**

*Evidence of electron-phonon coupling*

As already discussed, in these non-magnetic 2D materials, the phonon linewidths are determined by electron-phonon and phonon-phonon interactions. While the phonon-phonon interactions tend to broaden the linewidth as a function of temperature, electron-phonon interactions tend to reverse the trend [20,21]. Therefore, the effective thermal behaviour is determined by a competition between these two interactions. We have observed a suppression of the $\xi$-values as a function of the flake-thickness, which could be explained by an increase in the electron-phonon coupling with the addition of layers. Though the effect is vital for the explanation of the thermal behaviour as a function of flake-thickness, the electron-phonon interactions only showed an indirect effect on the discussed phonon linewidths as we did not observe any actual reversal of the temperature behaviour of the first-order modes for any of the flakes, and therefore, the contribution could only be understood by a decrease in the effective anharmonic behaviour as a function of the flake-thickness. However, as a further proof of the fact that the electron-phonon coupling increases with the increasing flake-thickness, we have observed the reversal of the thermal behaviour (direct evidence) of one of the second-order modes, $\omega_2$, which could be observed in $MoS_2$ using the 633 nm excitation laser source. Figure S18(a) compares the Raman spectra obtained at 300 K and 100 K for various flake-thicknesses of $MoS_2$. While it was not possible to systematically analyse the data of the ultrathin flakes of $MoS_2$ as a function of temperature because of the strong underlying photoluminescence feature as discussed previously, we could analyse the spectra obtained at 300 K and 100 K after suitable baseline corrections for all the flakes. The red (in 300 K spectra) and blue (in 100 K spectra) Lorentzian profiles represent the $\omega_2$ combinational mode. The corresponding linewidths (FWHMs) for the flakes of various thicknesses are shown in Figure S18(b). We observe that the linewidth of the $\omega_2$ mode shows a systematic decrease with increase in thickness. This is expected as the thinning down can substantially reduce the defects present in the flake [22]. Comparing the data obtained at 300 K and 100 K in Figure S18(b) (top panel), we observe that the linewidths of the $\omega_2$ mode for all the flakes show narrowing with increase in temperature, which is an indication of the electron-phonon coupling. The difference between the linewidths obtained at 100 K and 300 K is termed as $\Delta$(FWHM) in Figure S18(b) (bottom panel) that represents the extent of the anomaly (i.e. narrowing of FWHM with temperature), and hence, is proportional to the amount of the electron-phonon coupling. Interestingly, we observe that the anomaly increases as a function of the flake-thickness and is strongest for the bulk flake. Figure S18(c) shows the complete anomalous trend of the linewidth of the $\omega_2$ mode as a function of temperature for the 8L-$MoS_2$ flake, with the data for all the intermediate temperatures. These observations confirm the presence of electron-phonon coupling in all the flakes and the observed increase in the anomaly with the flake-thickness also confirms that the electron-phonon coupling increases as a function of the flake-thickness. It is important to note that the $\omega_2$ mode has been previously reported to be associated with electron-phonon coupling. Rao *et al.* [22] reported that this mode shows stronger electron-phonon coupling (manifested by a stronger temperature-dependence) in the bulk flakes, as opposed to the thinner nanoflakes they investigated. This is also consistent with the high $\chi$-value observed for the $\omega_2$ mode for 8L-flakes (Figure 4 in the main text).

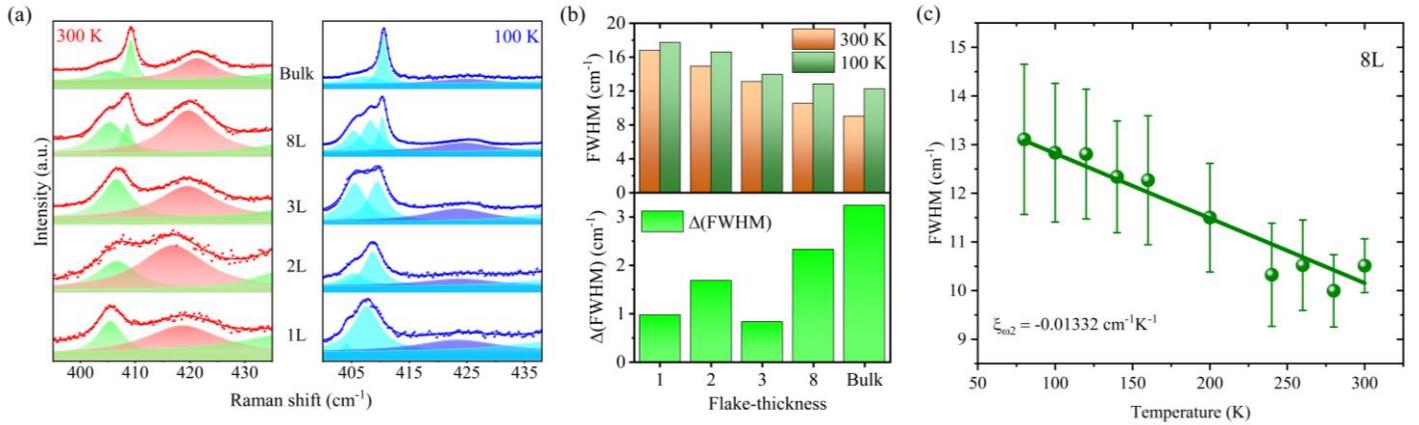

Figure S18. Evidence of electron-phonon coupling. (a) Comparison between the 300 K and 100 K Raman spectra of MoS$_2$ having various layer-thicknesses. The red (blue) spheres represent the data points, the red (blue) solid lines represent the overall fit, the red (blue) Lorentzian profiles represent the $\omega_2$ mode and the green (cyan) Lorentzian profiles represent the other modes in the 300 K (100 K) spectra; (b) The comparison between the linewidths (FWHMs) of the $\omega_2$ mode for all the investigated flakes reveal an anomalous broadening with decrease in temperature. The anomalous broadening, Δ(FWHM), which is a measure of the extent of electron-phonon coupling, shows a fairly systematic increase with flake-thickness; (c) The anomalous behaviour of the $\omega_2$ mode for 8L-MoS$_2$ reveals a negative slope due to the electron-phonon coupling associated with the mode.

**Supplemental note 5**

*Estimation of substrate-induced biaxial strain effect*

The thermal expansion coefficients of the investigated 2D flake and the underlying substrate shows a considerable mismatch as a function of temperature. This effectively results in an additional temperature-dependent biaxial strain on the 2D flake. The situation is explained schematically in Figure S19. Therefore, an accurate analysis of the thermal behaviour of the phonons must take account of this substrate-induced biaxial strain effect, which must be subtracted from the temperature-dependent response to obtain the actual temperature contribution to the phonon properties of the investigated flakes. Yoon *et al.* [23] have derived a very effective way to eliminate the effect of the strain induced due to the substrate from the temperature-dependent mismatch in the thermal expansion coefficients of the substrate and the investigated 2D flake. The frequency shift of a phonon [$\Delta\omega = \omega(T) - \omega_0$] may be written as follows:

$\Delta\omega = \Delta\omega_{actual} + \Delta\omega_{substrate}$

where, $\Delta\omega_{substrate} = \beta \int_{300\,K}^{T} (\alpha_{SiO_2}(T') - \alpha_{2D}(T'))\, dT'$

Here, $\beta$ is the biaxial strain coefficient, $\alpha_{SiO_2}(T')$ and $\alpha_{2D}(T')$ are the thermal expansion coefficients of SiO$_2$ and the investigated 2D flake at temperature $T'$. The biaxial strain coefficient $\beta$ can be obtained from uniaxial strain-dependent studies and is given by $\beta = -2\omega_0\gamma$ [23, 24], where $\gamma$ is the Grüneisen parameter of the corresponding phonon. The Grüneisen parameter for various modes may be obtained from uniaxial strain-dependent measurements. The application of uniaxial strain on a 2D flake results in a splitting of the $E_{2g}$-type in-plane modes, while they cause a redshift in frequency for $A_{1g}$-type out-of-plane modes. The corresponding Grüneisen parameters for the $E_{2g}$-type and $A_{1g}$-type modes are given by the following relations:

$$\gamma_{E_{2g}} = -\frac{\Delta\omega_{E_{2g}}^+ + \Delta\omega_{E_{2g}}^-}{2\omega_{E_{2g}}^0 (1-\nu)\varepsilon}$$

$$\gamma_{A_{1g}} = -\frac{\Delta\omega_{A_{1g}}}{\omega_{A_{1g}}^0 (1-\nu)\varepsilon}$$

where $\Delta\omega_{E_{2g}}^+$, $\Delta\omega_{E_{2g}}^-$, and $\Delta\omega_{A_{1g}}$ are the frequency redshifts of the two split-components of the $E_{2g}$ mode and the $A_{1g}$ mode, respectively, due to application of $\varepsilon$% of uniaxial strain. Conley *et al.* [25] reported $\gamma_{E_{2g}}$ for both monolayer and bilayer MoS$_2$ as ~1.06. Again, Lee *et al.* [26] reported a $\gamma_{E_{2g}}$ value of ~1 for bilayer MoS$_2$. Rice *et al.* [27] reported $\gamma_{E_{2g}}$ and $\gamma_{A_{1g}}$ for monolayer MoS$_2$ as ~0.65 and ~0.21, respectively. Based on various Raman studies on strain-dependence of MoS$_2$, we have chosen the Grüneisen parameter values for the $E_{2g}$ and $A_{1g}$ modes as ~1.06 and ~0.21, respectively, for our calculations. Using the $\gamma$ values as mentioned above and the thermal expansion coefficients of SiO$_2$ [23] and monolayer MoS$_2$ by Zhan-Yu *et al.* [28], we have estimated the substrate effect on the $E_{2g}$ and $A_{1g}$ phonons (F$_2$ and F$_3$) of monolayer MoS$_2$ (Figure 8 in main text). On the other hand, there are no reports on strain-dependent studies of monolayer MoTe$_2$. A recent study by Karki *et al.* [29] shows a very weak strain-dependence of the $E_{2g}$ mode for few-layer MoTe$_2$. As also discussed above for MoS$_2$, the Grüneisen parameter remains nearly constant for the ultrathin flakes of TMDs. Therefore, assuming the Grüneisen parameter for MoTe$_2$ to be independent of the flake-thickness, we have used the data by Karki *et al.* [29] for few-layer MoTe$_2$ to compute the $\gamma_{E_{2g}}$ of monolayer MoTe$_2$. We have used the Poisson's ratio value of ~0.25 for monolayer MoTe$_2$, as reported by Mortazavi *et al.* [30]. Using these data, we have been able to estimate the biaxial strain effect induced by the SiO$_2$/Si substrate on the thermal behaviour of the F$_3$ mode ($E_{2g}$) of MoTe$_2$, as shown in Figure S20. As opposed to MoS$_2$, we observe the substrate effect to be very weak in case of MoTe$_2$.

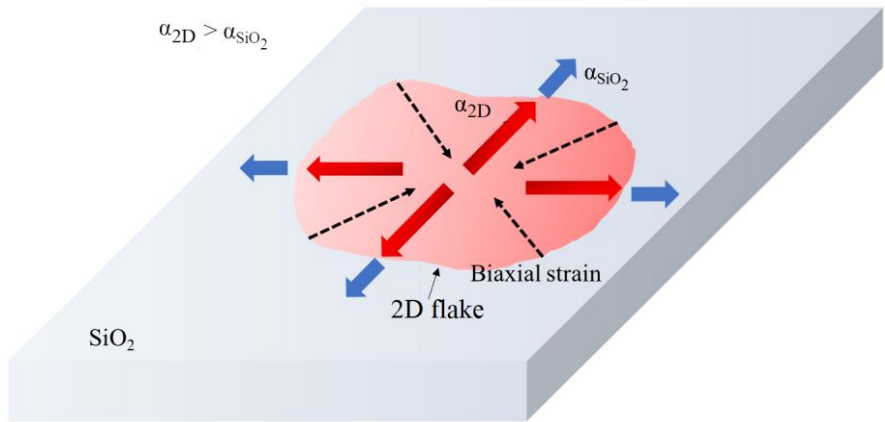

Figure S19. Schematic illustration of the biaxial strain induced due to mismatch in thermal expansion coefficient between the monolayer 2D flake and the underlying substrate. The blue and red arrows represent the different extents of thermal expansions undergone by the substrate and the 2D layer, respectively. The dashed black arrows represent the resultant compressive biaxial strain on the monolayer flake.

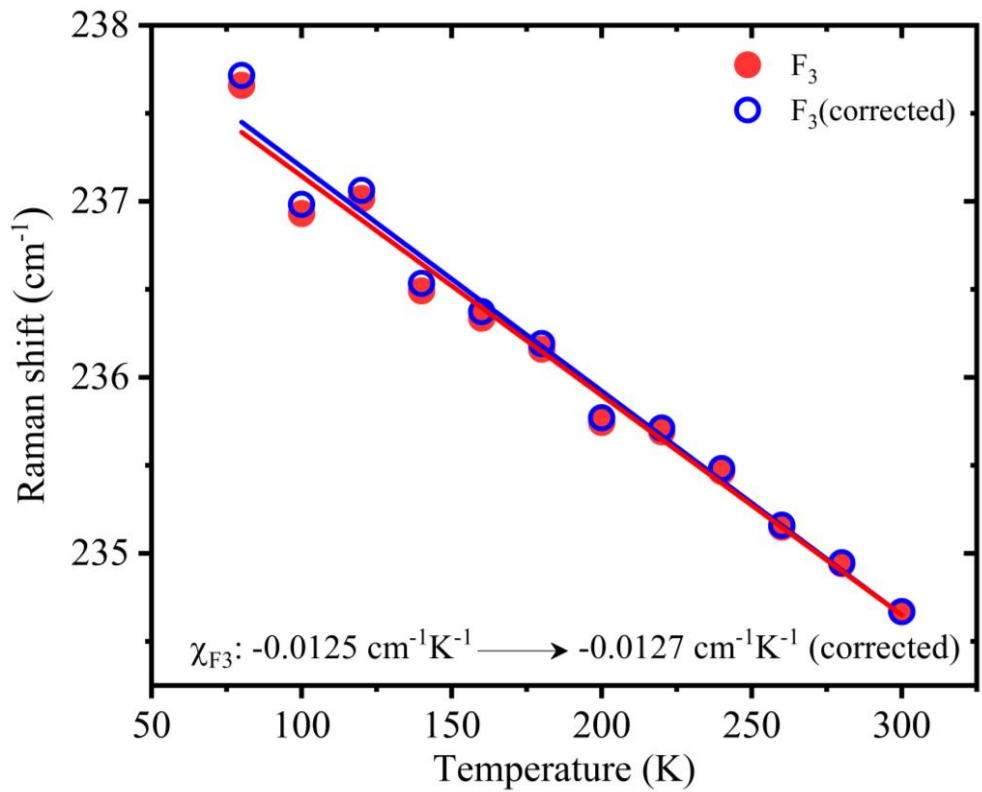

Figure S20. The temperature dependence of the $E_{2g}$ ($F_3$) phonons of 1L-MoTe$_2$ when supported on SiO$_2$/Si substrate (solid red circles). The phonon frequencies obtained after correcting for the substrate-effect are shown by the blue hollow spheres. The corresponding linear fits show minimal change in the first-order temperature coefficient, $\chi$

**Supplemental note 6**

*Polarization angle-dependent Raman intensity and selection rules*

The intensity of various Raman modes shows polarization angle dependence based on Raman selection rules. In order to explore the role of polarization angle on the Raman mode intensities, we have obtained the Raman spectra for various flakes of MoS$_2$ and MoTe$_2$ in the parallel (XX) and cross (XY) polarization configurations, as shown in Figures S22-S24. Conventionally these configurations are represented by the notations: $\bar{Z}(XX)Z$ and $\bar{Z}(XY)Z$, respectively. The Z and the $\bar{Z}$ represents the directions of propagation of the incident and scattered photons, respectively, XX represents the case when both the incident and scattered photons are polarized parallel to each other and XY represents the case when the polarizations are oriented perpendicular to each other. We clearly observe an almost complete suppression of the out-of-plane modes and the second-order modes in the XY configuration, while the in-plane modes show minimal changes. The polarization angle dependence of the mode intensities may be understood using the Raman selection rules, as follows.

The intensity of a Raman mode may be calculated by the following expression:

$$I \propto \sum_j \left| e_s^t . R_j . e_i \right|^2 \quad (A)$$

where $e_i$ and $e_s$ (the t superfix represents transpose of the matrix) are the unit vectors representing the polarization directions of the incident laser and the scattered Raman signal, and $R_j$ represents to the Raman tensor corresponding to the given mode. Depending on the degeneracy of the concerned mode, there may be multiple Raman tensors corresponding to a single mode. In such cases, the intensity of each of the tensors must be calculated using the expression above and the total intensity is given by the summation of the individual intensities.

Figure S21(a) shows the schematic for the setup we used to perform the polarization angle-dependent experiments. We have used a half waveplate in the path of the incident laser beam in order to rotate the polarization direction of the incident light with respect to the crystal axis of the sample. The polarization directions for both the incident and scattered light beams are initially oriented along the x-axis. The polarization direction of the incident light (white solid double-sided arrow in Figure S21(b)) is rotated by an angle θ with respect to the initial direction (along the x-axis) by rotating the half waveplate through an angle of θ/2. On the other hand, the polarization direction for the scattered light is kept fixed along the x-axis (yellow solid double-sided arrow in Figure S21(b)) throughout the entire experiment. Based on the above description of the system used, the unit vectors along the polarization direction of the incident and scattered light can be written as:

$$e_i = \begin{bmatrix} \cos\theta \\ \sin\theta \\ 0 \end{bmatrix}$$

$$e_s = \begin{bmatrix} 1 \\ 0 \\ 0 \end{bmatrix}$$

The Raman tensors corresponding to the phonons of symmetries $E_{1g}$, $E_{2g}$, and $A_{1g}$ are given as [31, 32]:

$$E_{1g} = \begin{bmatrix} 0 & 0 & 0 \\ 0 & 0 & c \\ 0 & c & 0 \end{bmatrix} \text{ and } \begin{bmatrix} 0 & 0 & -c \\ 0 & 0 & 0 \\ -c & 0 & 0 \end{bmatrix}$$

$$E_{2g} = \begin{bmatrix} 0 & d & 0 \\ d & 0 & 0 \\ 0 & 0 & 0 \end{bmatrix} \text{ and } \begin{bmatrix} d & 0 & 0 \\ 0 & -d & 0 \\ 0 & 0 & 0 \end{bmatrix}$$

$$A_{1g} = \begin{bmatrix} a & 0 & 0 \\ 0 & a & 0 \\ 0 & 0 & b \end{bmatrix}$$

Using the appropriate tensors in the expression (A), the intensity of the phonons of symmetries $E_{1g}$, $E_{2g}$, and $A_{1g}$ may be calculated as:

$$I_{E_{1g}} = 0$$

$$I_{E_{2g}} = |d|^2$$

$$I_{A_{1g}} = |a \cos(\theta)|^2$$

The polarization angle-dependent data is fitted using the above expressions in Figures 9 and 10 of the main text. The corresponding Raman spectra for the 2L-MoS$_2$ and 2L-MoTe$_2$ are shown in Figure S25-S27. The spectra clearly reveal the dependence of the different modes on the polarization angle, as discussed in the main text.

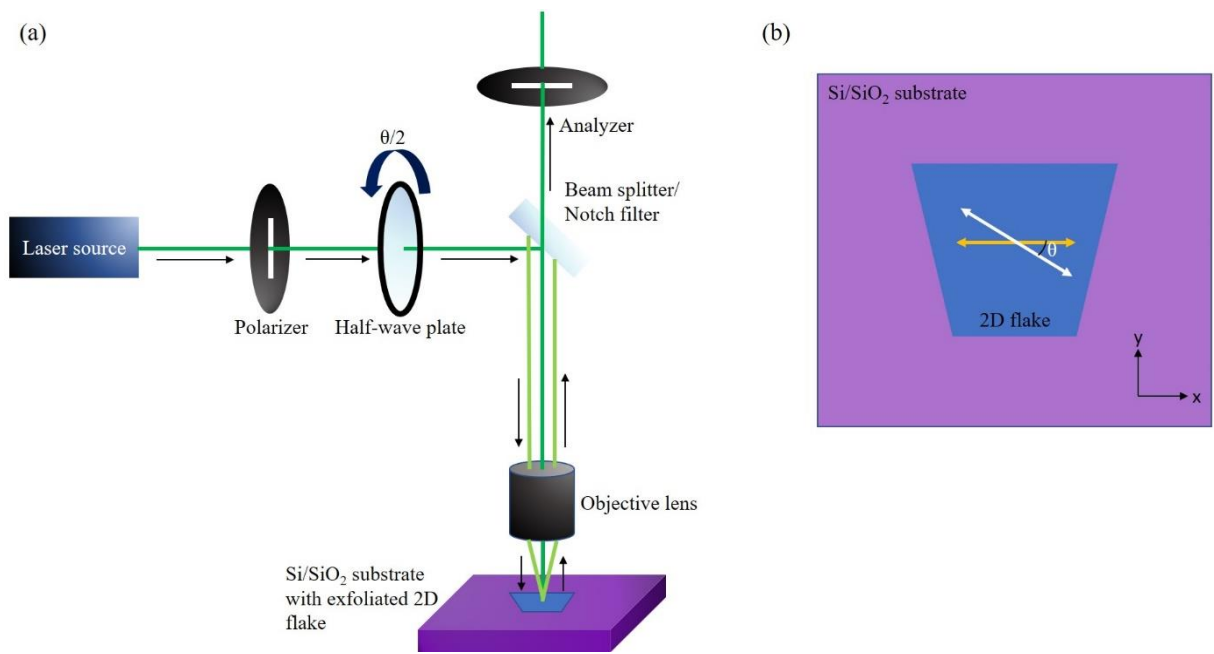

Figure S21. (a) The schematic of the system used to perform polarization angle-dependent Raman experiments on the MoS$_2$ and MoTe$_2$ flakes. The black arrows show the direction of propagation of the light beam. The incident laser beam travels from the laser source through a polarizer and then the polarized light is passed through a half waveplate, which rotates the polarization direction by an angle θ. The rotated beam then gets reflected by the beam splitter and is focussed on the sample through an objective lens. The scattered beam again passes through the objective and is transmitted through the beam splitter to the analyzer. (b) Top view of the sample with labelled polarization directions of the incident and scattered beams. The incident beam is rotated with respect to the default x-axis by an angle θ, while the scattered beam (direction of the analyzer) is oriented along the x-axis.

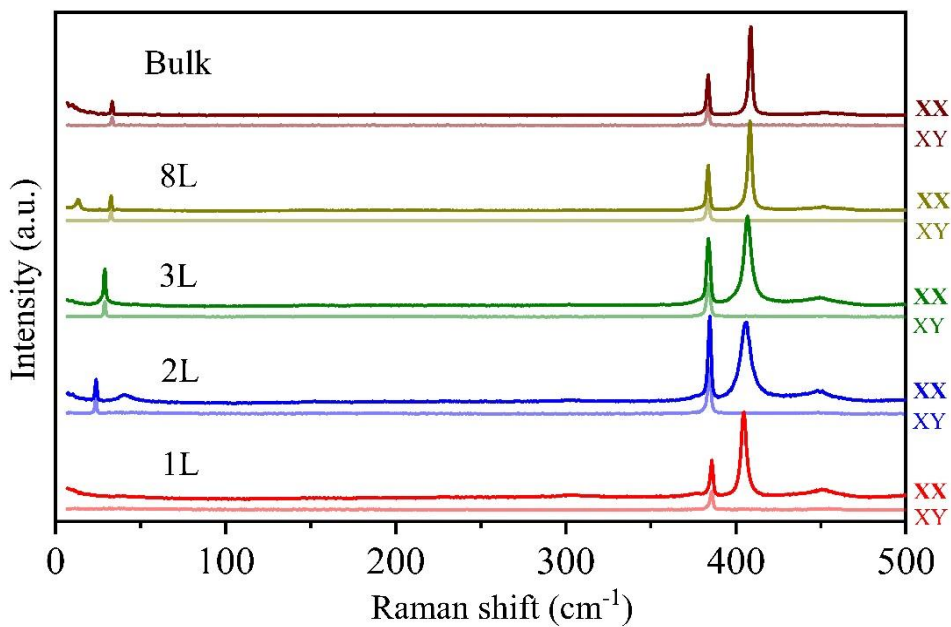

Figure S22. Raman spectra obtained for flakes of $MoS_2$ of various thicknesses in the parallel $\bar{Z}(XX)Z$ and cross $\bar{Z}(XY)Z$ polarization configurations using the 532 nm (non-resonant) laser excitation.

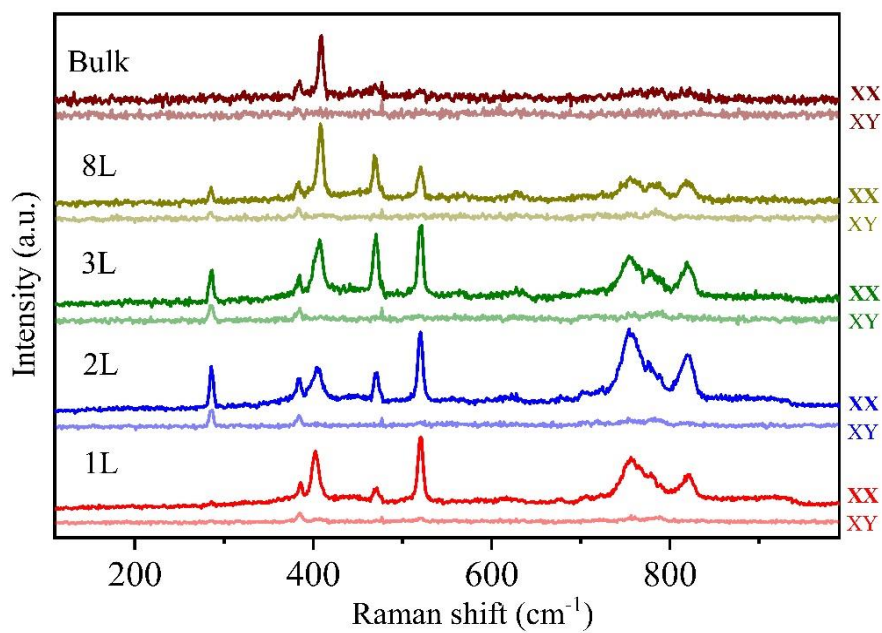

Figure S23. Raman spectra obtained for flakes of $MoS_2$ of various thicknesses in the parallel $\bar{Z}(XX)Z$ and cross $\bar{Z}(XY)Z$ polarization configurations using the 325 nm (resonant) laser excitation.

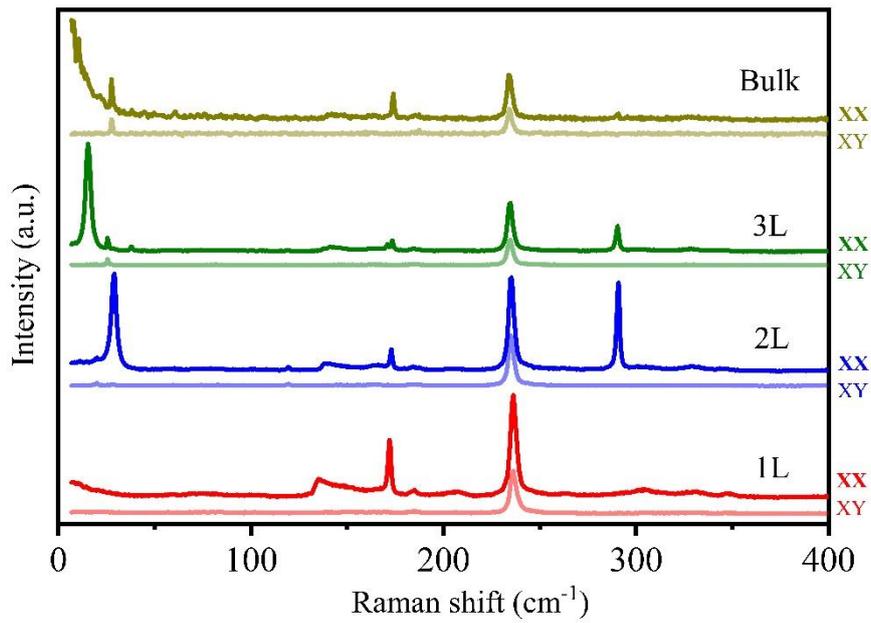

Figure S24. Raman spectra obtained for flakes of MoTe$_2$ of various thicknesses in the parallel $\bar{Z}(XX)Z$ and cross $\bar{Z}(XY)Z$ polarization configurations using the 532 nm (non-resonant) laser excitation.

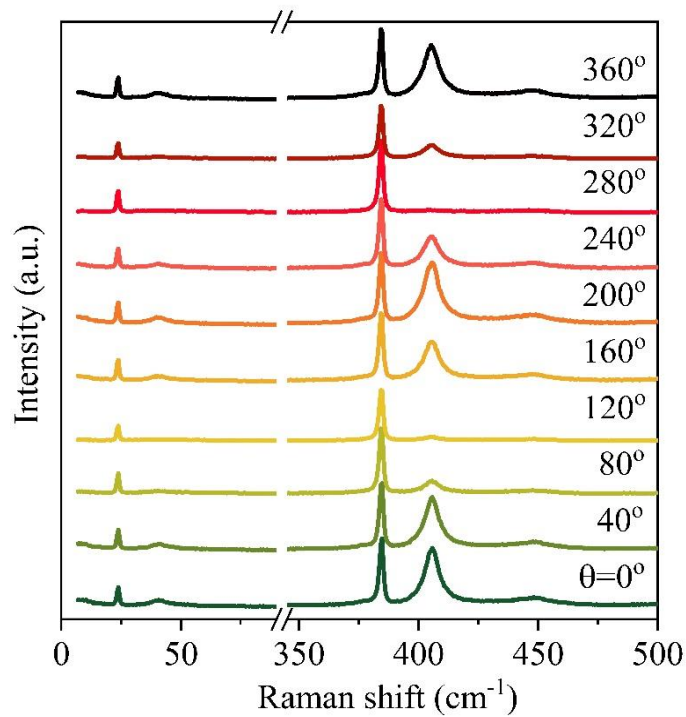

Figure S25. Raman spectra obtained for 2L-MoS$_2$ flake at different polarization angles (θ) using the 532 nm (non-resonant) laser excitation.

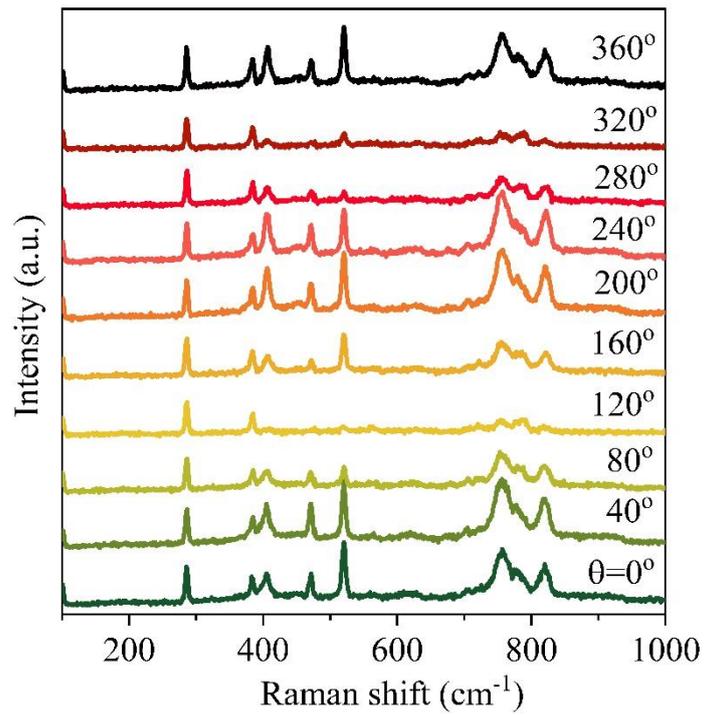

Figure S26. Raman spectra obtained for 2L-$MoS_2$ flake at different polarization angles (θ) using the 325 nm (resonant) laser excitation.

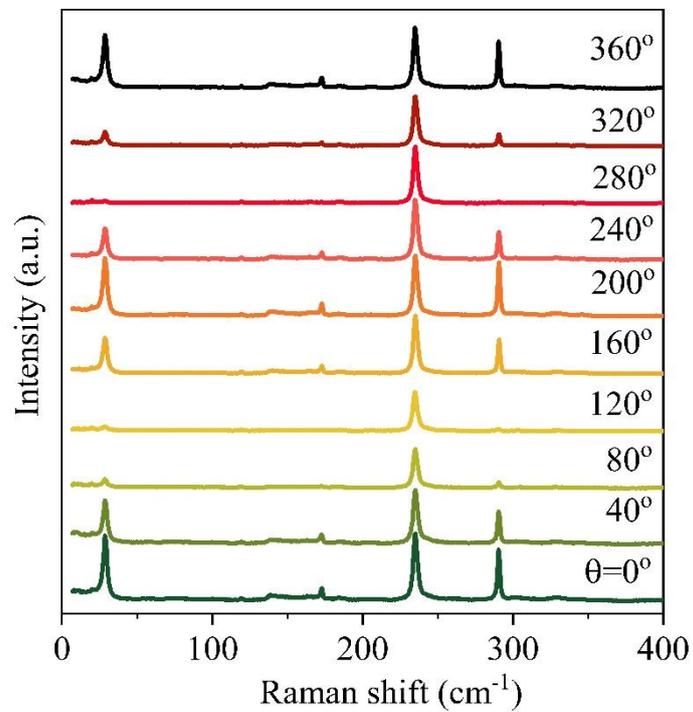

Figure S27. Raman spectra obtained for 2L-$MoTe_2$ flake at different polarization angles (θ) using the 532 nm (resonant) laser excitation.